\RequirePackage{rotating}
\documentclass[fleqn,usenatbib]{mnras}

\usepackage{newtxtext,newtxmath}
\usepackage[T1]{fontenc}
\usepackage{ae,aecompl}

\usepackage{caption}
\usepackage{graphicx}
\usepackage{longtable}
\usepackage{supertabular}
\usepackage{float}
\usepackage{rotating}%para usar sidewaystable

\usepackage{multicol}

\usepackage{color}

\usepackage{threeparttable}
\graphicspath{{./}}

\usepackage{amsmath}	% Advanced maths commands
\usepackage{amssymb}	% Extra maths symbols

\hypersetup{draft}

\title[NGC 4656 and its TDG candidate]{Kinematics and physical properties of the nearby galaxy NGC 4656 and its TDG candidate}

\author[N. Mu\~noz-Elgueta et al.]{
N. Mu\~noz-Elgueta,$^{1}$\thanks{E-mail: nahir@dfuls.cl}
S. Torres-Flores,$^{1}$
P. Amram,$^{2}$ 
J. A. Hernandez-Jimenez,$^{3}$
\newauthor F. Urrutia-Viscarra,$^{4}$
C. Mendes de Oliveira$^{3}$ and 
J.A. G\'omez-L\'opez$^{2}$
\\
$^{1}$Departamento de F\'\i{}sica y Astronom\'\i{}a, Universidad de La Serena, Av. Cisternas 1200, La Serena, Chile\\
$^{2}$Aix Marseille Univ., CNRS, CNES, LAM (Laboratoire d'Astrophysique de Marseille), Marseille, France\\
$^{3}$Instituto de Astronomia, Geof\'\i{}sica e Ci\^encias Atmosf\'ericas, Universidade de S\~ao Paulo, 05508-900 S\~ao Paulo, Brazil\\
$^{4}$Gemini Observatory/AURA, Southern Operations Center, Casilla 603 La Serena, Chile
}

\date{Accepted XXX. Received YYY; in original form ZZZ}

\pubyear{2018}

\begin{document}
\label{firstpage}
\pagerange{\pageref{firstpage}--\pageref{lastpage}}
\maketitle

\begin{abstract}
Interacting galaxies provide us with an excellent laboratory for studying a number of physical phenomena associated with these processes. In this paper, we present a spectroscopic and kinematic analysis of the interacting galaxy NGC 4656 and its companion Tidal Dwarf Galaxy (TDG) candidate, NGC 4656UV. Using Fabry-Perot  and GMOS multi-slit data, we investigated the possible origin of NGC 4656UV. We found that  NGC 4656UV has a low metallicity (12+log(O/H)$\sim$8.2) and it follows the mass-metallicity relation (MZR) for normal dwarf galaxies. For NGC 4656, we estimated a flat oxygen abundance gradient of $\beta$ = -0.027$\pm$0.029 dex kpc$^{-1}$, which suggests the presence of gas flows induced by gravitational interactions. By analysing radial velocity profiles and by fitting  a kinematic model of the observed velocity field, we confirm the literature result that NGC 4656 consists of one single body instead of two objects. We estimated a dynamical mass of $6.8^{1.8}_{-0.6}\times10^{9}$\,M$_{\sun}$ and R of 12.1\,kpc from the kinematic model of NGC 4656. Although the observed velocity field is dominated by rotation at large scales (V$_{max}$/$\sigma\gtrsim$2.8), important non-rotational motions are present at small scales. Based on these new results, and on previously published information, we propose that NGC 4656 and 4656UV are a pair of interacting galaxies.  NGC 4656UV is a companion of NGC 4656 and it does not have a tidal origin. The interaction between the two could have triggered the star formation in NGC 4656UV and increased the star formation in the northeast side of NGC 4656.
\end{abstract}

\begin{keywords}
galaxies: dwarf -- galaxies: individual (NGC 4656) -- galaxies: interactions -- galaxies: kinematics and dynamics
\end{keywords}

%%%%%%%%%%%%%%%%%%%%%%%%%%%%%%%%%%%%%%%%%%%%%%%%%%

%%%%%%%%%%%%%%%%% BODY OF PAPER %%%%%%%%%%%%%%%%%%

\section{Introduction}
The hierarchical scenario of galaxy formation  argues that the interactions and mergers play a fundamental role, and they are the key to the growth of the structures observed today \citep[e.g.][]{1972toomre,1993kauffmann_b}. % Furthermore, studying interacting galaxies in the local Universe can give us important clues about phenomena that were very common in the distant Universe.
Morphology, kinematic, metallicity and other physical properties are affected by interactions during galaxy evolution. For example, in the case of galaxy pairs, since \citep{1972toomre} we know that due to the intense tidal forces, interacting galaxies often show elongated and extended tidal tails, which are composed by stars and interstellar gas.
On the other hand, several authors have studied the oxygen  abundances  in the disks of galaxies with star formation and in tidal tails of interacting systems \citep[e.g.][]{2012rich,2014torres}, finding metallicity gradients  flatter than in isolated systems \citep[e.g.][]{1994zari}. This is consistent with results obtained in numerical simulations, which show that  gas flows induced by  interactions redistribute the gas in such a way that the original abundance gradients present in the galactic discs progressively flatten as the interaction/merger stage  progresses \citep{2010rupke_a,2012torrey}. Other most common indicators of on-going interactions, from the galaxies kinematics, are: highly disturbed velocity fields, presence of non-circular motions, double nuclei, double components in the emission profiles that trace the gas, the presence of an anomalous kinematic structures, misalignment of the position angles of the stellar and gaseous major axes, and a discordance between approaching and receding sides of the rotation curve \citep{2003amram}.\\%During interactions  tidal forces can compress the gas, and thus facilitate star formation processes, so studying interacting systems will allow us to determine star formation rates in these galaxies. Recent studies have revealed that close galaxy pairs present enhanced star formation rates (SFRs) \citep{2013patton,2010woods}, however objects already merged present neutral gas fractions similar to galaxies that have not been disturbed \citep{2015ellison} . \\
%Morphological and kinematical changes may happen as a result of interactions.
%In the case of galaxy pairs, since \citep{1972toomre}, we know that gravitational interaction between galaxies results in intense tidal forces that deform the galaxies and completely alter their morphology. As a result of this, interacting galaxies show elongated and extended tidal tails, which are composed by stars and interstellar gas. %In some cases these tidal tails present optical emission produced by newly star clusters; several studies have reported the formation of new  star forming regions located in tidal tails due to ejection of material during galaxy collisions and the onset of star formation \citep{2011mullan,2012torres,2014torres}. 
An important and (by now) well known end-result of tidal interactions is the possible formation of new stellar systems. For example, if the expelled material from interacting galaxies is gravitationally bound, the formation of a Tidal Dwarf Galaxy (hereafter TDG) is possible. First suggestions that new objects may be produced when galaxies collide remote from 1956, with \citeauthor{1956zwicky} and more recently by \cite{1996hibbard}. \cite{1998duc} showed that newly formed objects in interacting/merging galaxies have high metallicities because they formed from material already enriched by previous generations of stars in the disk of the galaxy.  TDGs are born in gaseous clouds/tails produced by interaction events, they are embedded in reservoirs of enriched cold gas and are expected to be active by forming stars. To be defined as TDGs, these newly formed objects must be self-graviting  \citep{1999duc,2011duc,2012duc}. Since TDGs are formed by gas ejected from the disc of gas-rich galaxies, this new class of objects may not contain a halo of dark matter, fact that fundamentally differentiates them from primordial dwarf galaxies, which have high M/L \citep{2010bournaud}.
The fate of TDGs in the context of galaxy evolution is not yet well understood. With time, TDGs coud be converted into satellite galaxies \citep{1996huns}, but it is not yet clear if the TDGs can survive the intense tidal field and the energetic feedback produced by the formation and stellar evolution within them \citep{2015ploe}.   The existence of evolved TDGs has been suggested by \cite{2014duc}. If these objects survive, they could be part of the population of dwarf galaxies. In order to understand TDGs formation and physical properties of progenitor galaxies, detailed studies of interacting systems exhibiting TDG candidates are requested, motivating the current study.\\
The system that is the subject of the present study consists of a galaxy, NGC 4656 and a TDG candidate called NGC 4656UV, which are part of a larger group where NGC 4631 and NGC 4656 are the main galaxies. NGC 4656 is classified as a SB(s)m pec galaxy located at a distance of 5.1 Mpc \citep[][ redshift-independent distance measured with Tully-Fisher]{2014sorce} and shows a disturbed morphology which displays obvious signs of gravitational interaction.  In literature, several authors have studied this system, however, the nature of NGC 4656 and the origin of NGC 4656UV are still unclear. %NGC 4656UV is an UV-bright object located at the northeast of a particularly curved area of NGC 4656 that has been reported by \citep{2012schech,2012demello}. This bright northeast end of this system has also been catalogued as NGC 4657, hence, some authors refer to NGC 4656 as NGC 4656/4657.
In order to continue the investigation on the most probable scenario for the formation and evolution of these objects, in this paper we have obtained Gemini GMOS spectroscopic data and Fabry-Perot data suitable for a thorough kinematic study of the system. 
In section \ref{sec:system_data} we present the system to study, the data description and reduction. The data analysis are presented in the sections \ref{sec:analysis1} and \ref{sec:analysis2}. In section \ref{sec:results} we present our results, continuing with discussion and conclusions which are presented in sections \ref{sec:discussion} and \ref{sec:summary}, respectively.

\section{The system and observational data}\label{sec:system_data}
\subsection{NGC 4656 and its TDG candidate, NGC 4656UV}

The galaxy pair NGC 4656/NGC 4631  has been studied by several authors and different scenarios have been proposed about its nature. The bright northeast end of NGC 4656 has also been cataloged as NGC 4657, hence, some authors refer to this system as NGC 4656/4657. A few authors \citep{1964vaucouleurs,1973nilson} consider that NGC 4657 is a separate entity from NGC 4656. \cite{1964vaucouleurs} proposed that the system could consist of two irregular magellanic interacting galaxies, possibly similar to the Antennae Galaxies. On the other hand, \cite{1967burbidge} found that NGC 4656 resembles structurally to MCG 12-7-28, concluding that it may be a barred spiral with a weak arm. Subsequently, \cite{1968roberts} shows the distribution and content of neutral hydrogen for the NGC 4656/4631 pair, finding a neutral atomic hydrogen bridge connecting NGC 4656 and NGC 4631, suggesting that the gas has been expelled by NGC 4631 because of its gravitational  interaction with NGC 4656. One decade later this bridge was resolved by \cite{1978weli} who, using 21 cm hydrogen line observations, found that the bridge is composed mainly by two filamentary structures. A different scenario was proposed by \cite{1973nilson}, who considered that NGC 4657 could be a separate object superimposed onto NGC 4656. However, by analyzing surface brightness profiles, \cite{1983stayton} concluded that NGC 4656 and NGC 4657 were part of the same galaxy, and found no conclusive optical indicators indicating an interaction with their neighbor NGC 4631. By studying the H{\sc i} gas, \cite{1994rand} note that NGC 4656 presents a structure that is disturbed and they concluded it is due to interaction effects. They suggested that NGC 4656 could be a system with two loosely wrapped tidal arms, or a ring galaxy almost edge-on.

It is only in the work of \cite{2012schech} where it was announced the discovery of a TDG candidate in the NGC 4656 system, which was denominated NGC 4656UV. Through evolutionary synthesis models, they found that NGC 4656UV presents a low metallicity ($\sim$10 times smaller than NGC 4656), and associated the origin of this TDG candidate to an encounter between NGC 4656 and NGC 4631 about 260-290 Myrs ago. \cite{2012demello} confirmed that this object could be a TDG candidate in finding 8 UV sources with ages less than 100. On the other hand, a giant stellar tidal stream was discovered in the halo of NGC 4631, a part of which extends between NGC 4631 and NGC 4656 \citep{2015martinez}. However, these authors discarded that the origin of this stream is due to some previous interaction between these two galaxies, mainly because the inclination of NGC 4631 and NGC 4656 with respect to the orbital plane makes tidal formation inefficient. They suggested that the streams around NGC 4631 are results of interactions between this galaxy and its dwarf satellites.

The main properties of NGC 4656 and NGC 4656UV are summarized in Table \ref{table:properties}.

\begin{table*}
      \caption[]{Main properties of the sample.}
\label{table:properties}
\begin{threeparttable}
\begin{tabular}{lccccccccc}
\hline\hline 
ID & $\alpha$(2000) & $\delta$(2000) & V$_{syst}$ & Distance & Size & M$_{stellar}$ & M$_{HI}$& M$_{dyn}$ & SFR\\
 & h:m:s & d:m:s  & km s$^{-1}$ & Mpc  & arcmin & M$_{\odot}$ & M$_{\odot}$ & M$_{\odot}$ & M$_{\odot}$ yr$^{-1}$\\ \hline \\
NGC 4656 & 12:43:57.73\tnote{\emph{a}} & +23:10:05.30\tnote{\emph{a}}  & 646\tnote{\emph{a}}  & 8.21\tnote{\emph{a}}  & 12.88\tnote{\emph{a}}  &  -&1.7*10$^{9}$ \tnote{\emph{c}}& 1.9*10$^{10}$ \tnote{\emph{b}} & 0.666\tnote{\emph{b}}   \\
TDG candidate & 12:44:14.79\tnote{\emph{b}} & +32:16:48.26\tnote{\emph{b}}  & 570\tnote{\emph{b}} & 8.21\tnote{\emph{a}}  & 4.93\tnote{\emph{b}}   & 9*10$^{7}$\tnote{\emph{b}} & 3.8*10$^{8}$\tnote{\emph{b}} & 1.6*10$^{9}$\tnote{\emph{b}}  & 0.027\tnote{\emph{b}} \\
 \hline
\end{tabular}
\begin{tablenotes}
\footnotesize
\item[\emph{a}]{Values were taken from NED.}
\item[\emph{b}]{Values were taken from \cite{2012schech}}
\item[\emph{c}]{Values were taken from \cite{1978weli} }
\end{tablenotes}
\end{threeparttable}

\end{table*}

%\section{The system and observational data}
\subsection{GMOS Data}

Spectroscopic observations were collected with the Gemini Multi-Object Spectrograph \citep[GMOS,][]{2004hook} mounted on the Gemini North telescope in multi-slit mode, during the nights of 2013 May 11 and June 1 under the science program GN-2013A-Q-87 (PI: F. Urrutia-Viscarra). Since one of our goals is to study the physical properties of the system based on strong-line methods, we used \textit{GALEX} images for the whole system to select the bluest star-forming regions in NGC 4656 and its TDG candidate. After selecting the regions, we proceeded to build the multi-slit masks.  For this purpose, four images of 300 seconds each were obtained in the g'-band filter for three different fields: one of them includes the whole TDG candidate and other two cover the northeast and the southwest regions of NGC 4656.
% Due to the area it occupies in the sky, to take the observations NGC 4656 was divided into two parts: Target 1 was called to the northeast part and Target 2 was called to the southwest part. \\

\begin{figure*}
\includegraphics[scale=0.4]{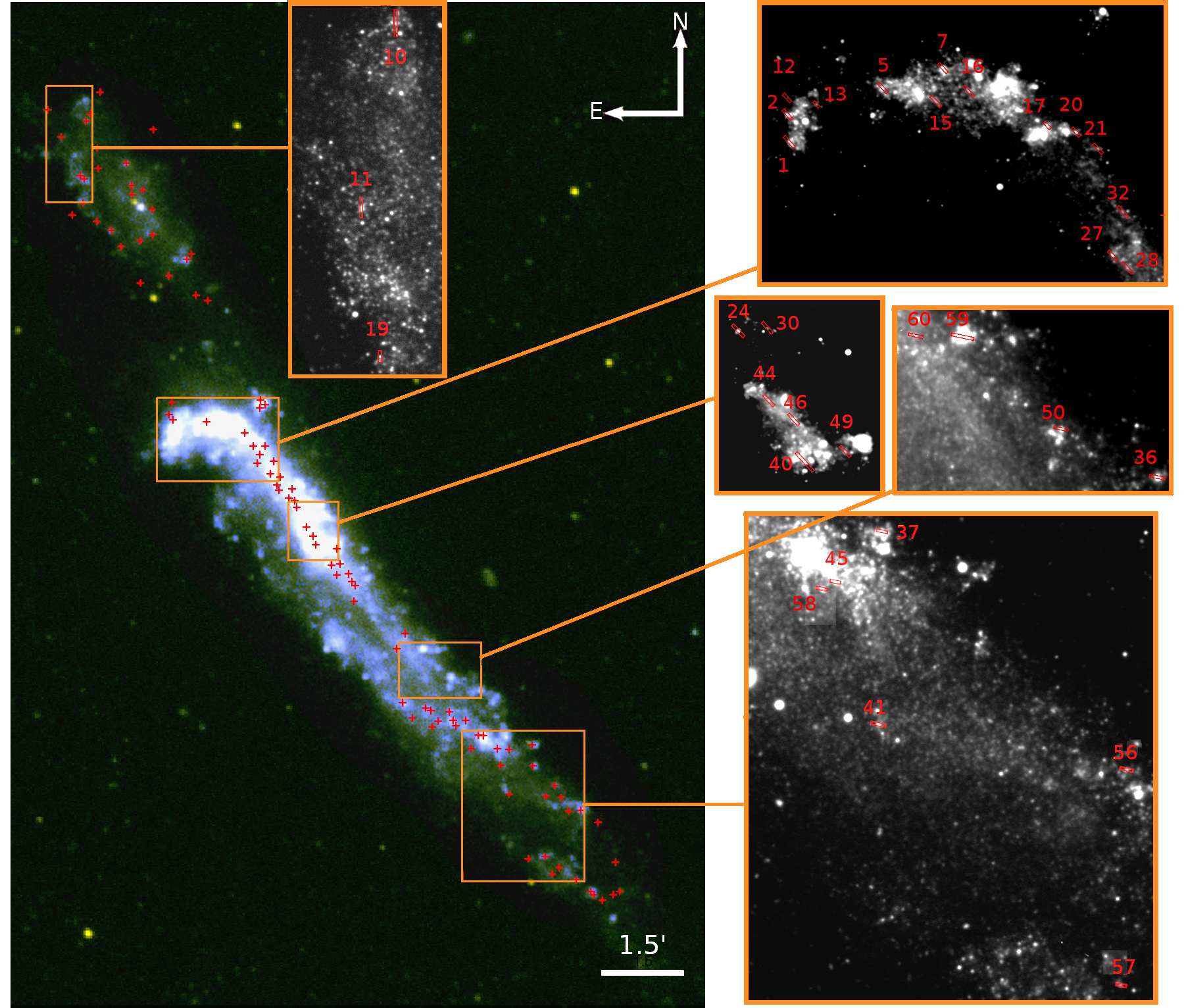}
\caption{False color image of NGC 4656 in the FUV and NUV bands (observed with \textit{GALEX}) where FUV represents the blue, the sum of FUV + NUV the green, and NUV the red. The red rectangles represent the slits used in GMOS observations, whose regions present spectra with emission lines. Each one is positioned with its respective orientation angle. The numbers on the slits indicate the identification of each region. The red crosses represent the GMOS regions which do not present H$\alpha$ emission.}
 \label{rgb}
%\caption{False color image of NGC 4656 in the FUV and NUV bands (observed with \textit{GALEX}) where FUV represents the blue, the sum of FUV + NUV the green, and NUV the red. The red rectangles represent the slits used in GMOS observations, whose regions present spectra with emission lines. Each one is positioned with its respective orientation angle. The numbers on the slits indicate the identification of each observed region.}
\end{figure*}

These images were processed with the Gemini iraf package (version 1.8), and the final combined images were used to build the multi-slit masks. Considering that our sources are mostly point-like, we adopted slit widths of 1 arcsec to maximize the incoming  light and to match the seeing. In order to improve the number of observed star-forming regions and the local sky sampling, the lengths of the slits were set for each object individually. The orientations of all slits are according to the instrumental position angle of 0$^{\circ}$ for TDG candidate, 43$^{\circ}$ for the field over NGC 4656 northeast (hereafter Target 1) and 258$^{\circ}$ for the field over NGC 4656 southwest (hereafter Target 2). In total, 52 regions with FUV emission were observed in Target 1, 48 in Target 2, and 30 in NGC 4656UV. Fig. \ref{rgb} shows a false color image of NGC 4656 and NGC 4656UV 
combined using the FUV, FUV+NUV and NUV filters from GALEX. Over this image red crosses have been superimposed representing the observed regions which do not present H$\alpha$ emission. %,  whose positions of are presented in the Appendix A, Table \ref{table:regions}. 
The orange boxes correspond to zooms of the GMOS images observed in the g'-band. The red rectangles with numbers represent the slits used, which contain only regions that have at least the H$\alpha$ emission line in their spectra, and that are at the redshift of the system (z$\sim$0.002). From here on, we will refer only to the spectra of these regions, since these were used to measure physical properties. The positions of these regions are presented in the Table \ref{table:flujos_posiciones}.

The observations were taken by using the grating R400$\_$G5305 and were performed taking exposures of 1570, 1300 and 1345 seconds for the TDG candidate, Target 1 and Target 2 respectively in three different central wavelengths:  6450, 6500 and 6550\AA\hspace{0.7mm} for each target. The spectral coverage ranged typically from 4700 to 7700\AA, which allowed us to measuring the emission lines necessary to estimate the physical properties of the system (H$\beta$, [O{\sc iii}]$\lambda\lambda$4959,5007\AA,  H$\alpha$, [N{\sc ii}]$\lambda\lambda$6548,6584\AA, [S{\sc ii}]$\lambda\lambda$6716,6731\AA). The spectral resolution at H$\alpha$ was R$\sim$875 ($\Delta\lambda\sim$7.5\AA).  Flat-field frames were taken after each science observation in order to avoid any effect caused by the instrumental flexure, and CuAr arc lamps spectra were observed at the end of each night.%    to proceed to flux transmission correction, spectral calibrations and to account for possible instrumental flexures.\\  

%\begin{sidewaystable*}\captionsetup{font=scriptsize}\caption{Posiciones, flujos y exceso de color de las regiones de formación estelar en NGC 4656 y NGC 4656UV}

\subsubsection{Data reduction and flux calibration}
The GMOS data were reduced using the {\sc Gemini} package version 1.13.1 in {\sc iraf}. All frames were bias subtracted,  trimmed,  and flat-fielded  using {\sc gbias}, {\sc gsflat} and {\sc gsreduce}. Cosmic  rays  were  removed  using {\sc lacos$\_$spec} \citep{2001vandokkum}. Each individual 2D spectrum, corresponding to one slit  for each central wavelength observed, was  cut  from  the  MOS  field using the task {\sc gscut}. All spectra were wavelength calibrated using the tasks {\sc gswavelength} and {\sc gstransform}. The rms value  of  the  wavelength  calibration was $\sim$0.3\AA\hspace{07mm} for each target. The combination of the three spectra taken at slightly different central wavelengths was done in order to remove  the  detector  gaps  and  the cosmic  rays  that were still present, using the task {\sc gcombine}.   The sky lines  were  removed with the task {\sc gsskysub}. Then, with the purpose to flux calibrate all spectra, similar procedures were applied to the spectrum of the standard star Wolf 1346. The sensitivity  function  was obtained with the task {\sc sensfunc}. The standard star Wolf 1346 was observed on a  different night from that of the main fields, on June 2, 2013. Finally, we obtained calibrated unidimensional spectrum for each region observed in NGC 4656 and NGC 4656UV.

%\centering

\subsection{Fabry-Perot Data}
The two-dimensional kinematic observations of NGC 4656 and the TDG candidate were carried out  March 2 to 6, 2016, using the GHASP instrument that consists of a focal reducer containing a Fabry-Perot interferometer, interference filters and an imaging photon counting system (IPCS). GHASP is mounted on the Cassegrain focus  of the  1.93  m  telescope  at  the  Observatoire  de Haute Provence (OHP). The field of view (FoV) of the instrument is $\sim$ 5.8 $\times$ 5.8 arcmin$^{2}$ and the IPCS detector \citep{2002gach} provides a pixel size of 0.68 arcsec pix$^{-1}$. The interference order of the Fabry-Perot (FP) used was p=798 at H$\alpha$, where the free spectral range (FSR) was 376 km s$^{-1}$. The Finesse of the interferometer is F=12, and in order to sample correctly the line, we scanned the FSR through 32 channels, hence the sampling step is about 12 km s$^{-1}$. A total exposure time of 10.4 hours was distributed over 11 overlapping field-of-views mapping the whole system, two of them corresponding to the TDG candidate and nine to NGC 4656. In order to minimize sky variation during the scanning process, thanks to the IPCS that is readout noise free,  each individual channel was observed with an exposure time of 10 seconds. The total observing time per field range between 1920 and 8940 seconds (see Table \ref{table:fp}).
 \begin{table}   
\centering
%\begin{center}
%\resizebox{0.35\textwidth}{!}{
\caption{Central coordinates and exposure times of the observed FP fields.}\label{table:fp}
%\resizebox{0.45\textwidth}{!}{
\begin{threeparttable}
 \begin{tabular}{lccc}
%\caption[]{Propiedades principales de la muestra}
\hline\hline 
ID\tnote{\emph{a}}& RA\tnote{\emph{b}} & Dec\tnote{\emph{b}} & t$_{exp}$\tnote{\emph{c}} \\ 
&J2000. &J2000. &s\\ \hline \\
1  &12:44:18 &+32:16:50 &8965    \\
2  &12:44:14 &+32:14:40 &3780  \\   
3  &12:44:14 &+32:13:20 &1920  \\   
4  &12:44:10 &+32:11:50 &1980    \\   
5  &12:44:06 &+32:10:40 &1980    \\   
6  &12:44:03 &+32:09:50 &1980    \\   
7  &12:43:59 &+32:08:40 &1980    \\   
8  &12:43:55 &+32:07:30 &1980    \\   
9  &12:43:50 &+32:06:30 &1980    \\   
10 &12:43:43 &+32:04:30 &3480    \\   
11 &12:43:34 &+32:03:10 &7380    \\     \hline
\end{tabular}
\begin{tablenotes}
\scriptsize
\item[\emph{a}]{Identification number for each observed field (north-south direction).}
\item[\emph{b}]{Central coordinates of the field, in J2000 epoch.}
\item[\emph{c}]{Total exposure time for each field, in seconds. }
\end{tablenotes} 
%\end{threeparttable}
%\end{center}
\end{threeparttable}
%}
\end{table}
\subsubsection{Data reduction}

The Fabry-Perot data were reduced by using the package developed by \cite{2006daigle}. This reduction package provides spatial adaptive binning,  based  on  the  2D  Voronoi  tessellation method, applied to the 3D data cubes. The spatial adaptive binning allows to keep the highest spatial resolution in bright H{\sc ii} regions meanwhile it increases the signal-to-noise ratio (SNR) in low flux areas in decreasing the spatial resolution. Using this  technique, bins are formed by adding new pixels until the region reaches a given level that is set a prior; this level is known as the signal-to-noise target (SNRt). For NGC 4656, we set SNRt=5 per bin. Nevertheless, the TDG candidate displays such a low H$\alpha$ emission over all its extension that the adaptative binning used for NGC 4656 is not suitable for the TDG candidate; thus we applied a Gaussian smoothing with a FWHM of 3 pixels. Astrometry was provided by using the {\sc koords} routine of the {\sc karma} package \citep{1996gooch}. 

Sky emission was removed through the subtraction of a sky cube, built from regions with no emission from the associated galaxies.
%To remove the sky emission, one must identify the sky-dominated regions present in the data cube. After that, a sky-cube is constructed and the OH sky lines are extracted. 
Once the reduction process is finished, we obtain the H$\alpha$ monochromatic map, the velocity  field and the velocity dispersion map for each field observed. These maps were finally cleaned through the routine {\sc clean$\_$maps}, where the low intensity emission associated with instrumental noise  was subtracted.

For each field, velocity dispersion maps obtained have been corrected from instrumental broadening. In this way, the real velocity dispersions $\sigma$ can be estimated through $\sigma$=$\sqrt[]{\sigma^{2}_{observed}-\sigma^{2}_{inst}}$, assuming that the observed and instrumental profiles can be fitted by Gaussian functions. The instrumental broadening was estimated from the average of dispersion maps derived from the lines of the Neon calibration lamp, whose resulting value was $\sigma_{inst}$=11.5$\pm$5.0 km s$^{-1}$,
where the uncertainty is represented by the standard deviation.

\subsubsection{Flux calibration}\label{datos_fluxcalibration}

H$\alpha$ monochromatic maps were flux-calibrated by using the Gemini GMOS spectra. %, similar to the procedure performed by \cite{2017mendes}. 
With this purpose, we have used the  H$\alpha$  flux of a total of 34 sources (where 3 sources are located in the TDG candidate, and 31 in NGC 4656). On the other hand, we have used the monochromatic Fabry-Perot maps to estimate the H$\alpha$ emission (in counts s$^{-1}$) in the same extraction windows used for the Gemini data. Then, we compared the H$\alpha$ fluxes (in erg s$^{-1}$cm$^{-2}$) with the H$\alpha$ emission measured for the Fabry-Perot data by fitting a linear fit on the data and fixing the zero point equal to zero. The final fit provided us a coefficient of 2.19$\times$10$^{-17}$erg count$^{-1}$cm$^{-2}$ which was used to calibrate the H$\alpha$ monochromatic maps for NGC 4656 and the TDG candidate. The root mean square error (RMSE) of the linear relation is 1.32$\times$10$^{-15}$erg s$^{-1}$cm$^{-2}$.

%%%%%%%%%%%%%%%%%ANALYSIS%%%%%%%%%%%%%%%%%%%%%%%%%%%%%%%%%%%%%%%%
%%%%%%%%%%%%%%%%%%%%%%%%%%%%%%%%%%%%%%%%%%%%%%%%%%%%%%%%%%%%%%%%%%%%%%%%%%%%%%

\section{Spectroscopy: physical parameters}\label{sec:analysis1}

\subsection{Extinction}
The spectra of the regions observed with GMOS show negligible stellar continuum compared with the strong emission from the nebular lines. Since the observed emission is affected by Galactic and internal extinction, it is necessary to make corrections. The observed fluxes were corrected for Galactic extinction through the {\sc idl} code {\sc fm$\_$unred} \citep{1999fitz}, using a colour excess of E(B-V)=0.011 (NED database). There is a different internal extinction for each region, since it depends on the internal properties of each source. The spectral coverage of our observations allows us to derive the H$\alpha$/H$\beta$ ratio to determine the internal extinction through the Balmer decrement. We have computed intrinsec H$\alpha$/H$\beta$ ratio in referring to \cite{1989oster}, considering a typical electronic temperature
and density for H{\sc ii} regions of T$_{e}$=10000 K and N$_{e}$=100. We consider hereafter this value as a lower threshold. Then, the observed fluxes for all spectra were corrected for internal reddening using the {\sc idl} code {\sc calz$\_$unred} \citep{2000calzetti}, which assumes a starburst extinction law. We have first estimated the nebular colour excess following the recipes given in \cite{2013dominguez}, and then we obtained the stellar colour excess that in this case is expressed by E(B-V)$_{star}$=0.44$\times$E(B-V)$_{gas}$. The sources for which H$\beta$ emission was not detected (regions $\#$10, 11, 12, 44, 45, 50, 56, 57, 58 and 60) and those whose for which H$\alpha$/H$\beta$ ratio was lower than the intrinsic value defined above (regions $\#$1, 40, 13 and 37, with emission lines contaminated by noise) were not corrected by internal extinction because it was not possible to estimate E(B-V)$_{star}$. The values of the color excesses, for the regions in which this measurement was possible, are listed in the third column of Table \ref{table:flujos_posiciones}.

\subsection{Emission lines measurements}

Once the spectra were corrected by extinction, we measured the nebular emission-line fluxes corresponding to H$\beta$, [O{\sc iii}]$\lambda\lambda$4959,5007\AA,  H$\alpha$, [N{\sc ii}]$\lambda\lambda$6548,6584\AA\hspace{0.7mm} and [S{\sc ii}]$\lambda\lambda$6716,6731\AA. We have used the {\sc splot} task in the {\sc noao} package from {\sc iraf}, fitting Gaussian models to the emission lines. The {\sc splot} task provides uncertainties for the fluxes measurements, which are computed by Monte-Carlo simulation. In Table \ref{table:flujos_posiciones} we tabulate the fluxes and their uncertainties for each region observed with GMOS.

\begin{sidewaystable*}\caption{Positions, fluxes and color excesses for star-forming regions in NGC 4656 and NGC 4656UV}
%\begin{table}[ht!] \captionsetup{font=scriptsize}\caption{Posiciones y flujos de las regiones de formación estelar en NGC 4656 y NGC 4656UV}
\label{table:flujos_posiciones}
%\begin{center}
\centering 
\resizebox{\textwidth}{!}{
\begin{threeparttable}
\begin{tabular}{lccccccccccc}
\hline\hline  
ID     &R.A.           &Dec.        &E(B-V)$_{est}$\tnote{\emph{a}} &H$\beta$   &[O{\sc iii}]$\lambda$4959  &[O{\sc iii}]$\lambda$5007 &H$\alpha$  &[N{\sc ii}]$\lambda$6548  &[N{\sc ii}]$\lambda$6584  &[S{\sc ii}]$\lambda$6716  &[S{\sc ii}]$\lambda$6731 \\
& J2000. & J2000.& & & &$\times$10$^{-15}$erg&cm$^{-2}$s$^{-1}$ & & & &\\ \hline \\ 	  
	  
  1      &12:44:11.513 &+32:12:14.37& -    	&0.059 $\pm$0.004  &0.040    $\pm$0.004 &0.076    $\pm$0.006 &0.141   $\pm$0.004   &0.004   $\pm$0.006 &0.007   $\pm$0.006 &0.012 $\pm$0.005 &0.008 $\pm$0.007 	\\%ap2	  
% 1(ap2) &12:44:11.513 &+32:12:14.37&     	&      -           &              -     &              -     &0.140   $\pm$0.001   &-       -          &             -     &           -     &          -     	\\  	  
  2      &12:44:11.564 &+32:12:23.65& 0.073     &1.900 $\pm$0.028  &1.890    $\pm$0.028 &5.630    $\pm$0.032 &6.330   $\pm$0.029   &0.066   $\pm$0.052 &0.224   $\pm$0.036 &0.265 $\pm$0.033 &0.192 $\pm$0.035	\\	  
  5      &12:44:09.003 &+32:12:32.76& 0.050     &1.600 $\pm$0.029  &1.660    $\pm$0.028 &4.890    $\pm$0.025 &4.940   $\pm$0.024   &0.038   $\pm$0.050 &0.131   $\pm$0.034 &0.321 $\pm$0.027 &0.244 $\pm$0.033	\\	  
  7      &12:44:07.389 &+32:12:39.39& 0.134     &1.380 $\pm$0.008  &1.980    $\pm$0.008 &5.800    $\pm$0.008 &4.690   $\pm$0.008   &0.066   $\pm$0.010 &0.134   $\pm$0.009 &0.223 $\pm$0.009 &0.158 $\pm$0.010	\\
  10     &12:44:19.416 &+32:18:44.89& -   	&     -            &        -           &        -           &0.043   $\pm$0.004   &       -           &      -            &     -           &     -         	\\%original id 5	
   11    &12:44:20.343 &+32:17:41.99& -   	&     -            &        -           &        -           &0.207   $\pm$0.009   &       -           &      -            &0.012 $\pm$0.013 &0.008 $\pm$0.020	\\	  
  12     &12:44:11.568 &+32:12:29.31& -   	&     -            &        -           &        -           &0.090   $\pm$0.007   &       -           &       -           &0.014 $\pm$0.011 &     -          	\\	  
  13     &12:44:10.788 &+32:12:27.18& -   	&0.167 $\pm$0.015  &        -           &        -           &0.329   $\pm$0.006   &       -           &       -           &     -           &     -          	\\	  
  15     &12:44:07.575 &+32:12:28.16& 0.034     &0.216 $\pm$0.018  &0.246    $\pm$0.019 &0.482    $\pm$0.016 &0.598   $\pm$0.016   &0.015   $\pm$0.035 &0.029   $\pm$0.021 &0.047 $\pm$0.019 &0.036 $\pm$0.019	\\%ap1	  
  16     &12:44:06.692 &+32:12:31.58& 0.007     &0.590 $\pm$0.021  &0.319    $\pm$0.019 &1.030    $\pm$0.020 &1.720   $\pm$0.019   &0.021   $\pm$0.024 &0.063   $\pm$0.021 &0.117 $\pm$0.022 &0.084 $\pm$0.021	\\	  
  17     &12:44:04.581 &+32:12:20.01& 0.052     &0.678 $\pm$0.027  &1.020    $\pm$0.025 &2.920    $\pm$0.026 &2.090   $\pm$0.027   &0.041   $\pm$0.140 &0.035   $\pm$0.030 &0.085 $\pm$0.030 &0.053 $\pm$0.028	\\	
  19     &12:44:19.845 &+32:16:51.42& 0.119     &0.257 $\pm$0.018  &0.289    $\pm$0.015 &0.826    $\pm$0.015 &0.917   $\pm$0.016   &-       -          &       -           &0.036 $\pm$0.019 &0.028 $\pm$0.021	\\	    
  20     &12:44:03.812 &+32:12:17.74& 0.178     &0.724 $\pm$0.044  &0.947    $\pm$0.044 &2.950    $\pm$0.042 &2.880   $\pm$0.038   &-       -          &0.115   $\pm$0.063 &0.165 $\pm$0.063 &0.140 $\pm$0.062	\\	  
  21     &12:44:03.225 &+32:12:12.10& 0.064     &1.520 $\pm$0.026  &1.590    $\pm$0.031 &4.750    $\pm$0.029 &4.780   $\pm$0.028   &0.037   $\pm$0.040 &0.161   $\pm$0.030 &0.245 $\pm$0.029 &0.175 $\pm$0.031	\\	  
  24     &12:43:59.918 &+32:10:51.67& 0.031     &4.880 $\pm$0.031  &6.330    $\pm$0.030 &19.100   $\pm$0.030 &14.200  $\pm$0.029   &0.154   $\pm$0.036 &0.523   $\pm$0.033 &0.580 $\pm$0.032 &0.455 $\pm$0.033	\\	  
  27     &12:44:02.830 &+32:11:35.80& 0.149     &1.230 $\pm$0.033  &2.020    $\pm$0.033 &5.850    $\pm$0.031 &4.480   $\pm$0.030   &0.043   $\pm$0.040 &0.136   $\pm$0.038 &0.165 $\pm$0.028 &0.143 $\pm$0.032	\\	  
  28     &12:44:02.435 &+32:11:31.49& 0.098     &0.493 $\pm$0.023  &0.793    $\pm$0.024 &2.340    $\pm$0.025 &1.810   $\pm$0.024   &0.022   $\pm$0.049 &0.075   $\pm$0.025 &0.210 $\pm$0.025 &0.147 $\pm$0.023	\\%ap2	  
  30     &12:43:59.136 &+32:10:52.71& 0.052     &1.210 $\pm$0.027  &0.906    $\pm$0.027 &2.620    $\pm$0.028 &3.830   $\pm$0.026   &0.090   $\pm$0.026 &0.244   $\pm$0.026 &0.341 $\pm$0.028 &0.235 $\pm$0.024	\\	  
  32     &12:44:02.555 &+32:11:50.53& 0.056     &0.684 $\pm$0.019  &0.754    $\pm$0.021 &2.110    $\pm$0.021 &2.230   $\pm$0.022   &0.039   $\pm$0.031 &0.083   $\pm$0.026 &0.119 $\pm$0.021 &0.090 $\pm$0.021	\\	
  36     &12:43:43.583 &+32:07:24.89& 0.062     &0.135 $\pm$0.036  &0.221    $\pm$0.036 &0.585    $\pm$0.037 &0.463   $\pm$0.027   &        -          &       -           &0.027 $\pm$0.110 &0.017 $\pm$1.300	\\
  37     &12:43:41.297 &+32:06:36.43& -   	&0.161 $\pm$0.009  &0.052    $\pm$0.006 &0.171    $\pm$0.007 &0.406   $\pm$0.007   &0.006   $\pm$0.027 &0.005   $\pm$0.007 &0.023 $\pm$0.008 &0.012 $\pm$0.008	\\
  40     &12:43:58.130 &+32:10:07.00& -   	&0.295 $\pm$0.019  &0.343    $\pm$0.018 &1.230    $\pm$0.017 &0.795   $\pm$0.020   &        -          &0.047   $\pm$0.021 &0.076 $\pm$0.023 &0.054 $\pm$0.024	\\%ap2
  41     &12:43:41.396 &+32:05:30.73& 0.049     &0.239 $\pm$0.011  &0.179    $\pm$0.011 &0.440    $\pm$0.011 &0.744   $\pm$0.011   &        -          &0.014   $\pm$0.020 &0.023 $\pm$0.012 &0.018 $\pm$0.015	\\
  44     &12:43:59.098 &+32:10:27.88& -         &      -           &        -           &0.044    $\pm$0.006 &0.073   $\pm$0.004   &        -          &0.010   $\pm$0.008 &0.016 $\pm$0.0056&0.014 $\pm$0.009	\\  %ap1
  45     &12:43:42.553 &+32:06:19.29& -   	&      -           &          -         &         -          &0.049   $\pm$0.004   &        -          &       -           &      -          &     -    	\\
  46     &12:43:58.432 &+32:10:21.47& 0.059     &2.070 $\pm$0.025  &4.160    $\pm$0.028 &12.700   $\pm$0.025 &6.360   $\pm$0.024   &0.044   $\pm$0.058 &0.085   $\pm$0.029 &0.190 $\pm$0.028 &0.148 $\pm$0.031	\\ %ap1	  
 %46     &12:43:58.432 &+32:10:21.47& 0.14	&1.680 $\pm$0.059  &2.490    $\pm$0.049 &7.690    $\pm$0.048 &12.000  $\pm$0.047   &0.118   $\pm$0.059 &0.436   $\pm$0.069 &0.807 $\pm$0.064 &0.710 $\pm$0.057	\\ %ap2
  49     &12:43:57.048 &+32:10:10.56& 0.138     &1.810 $\pm$0.047  &3.010    $\pm$0.043 &8.870    $\pm$0.047 &6.180   $\pm$0.042   &0.068   $\pm$0.042 &0.130   $\pm$0.055 &0.234 $\pm$0.051 &0.153 $\pm$0.040	\\%ap2	  
  50     &12:43:46.162 &+32:07:41.30& -   	&                  &           -        &         -          &0.043   $\pm$0.008   &        -          &             -     &           -     &     -          	\\	 
  56     &12:43:34.739 &+32:05:15.36& -   	&     -            &        -           &         -          &0.145   $\pm$0.007   &        -          &              -    &           -     &     -          	\\%origi id 16	  
  57     &12:43:34.897 &+32:04:01.92& -   	&     -            &        -           &         -          &0.020   $\pm$0.008   &        -          &              -    &           -     &     -          	\\%ori id 17                                                                                                                                                                                                                   
  58     &12:43:42.910 &+32:06:16.89& -   	&     -            &           -        &         -          &0.079   $\pm$0.010   &        -          &             -     &           -     &     -          	\\%original id 46                                                                                                                                                                                                                   
  59     &12:43:48.812 &+32:08:12.76& 0.041     &7.480 $\pm$0.050  &15.700   $\pm$0.045 &49.100   $\pm$0.048 &21.800  $\pm$0.048   &0.082   $\pm$0.076 &0.255   $\pm$0.052 &0.593 $\pm$0.053 &0.410 $\pm$0.051	\\	  
  60     &12:43:50.069 &+32:08:13.23& -   	&     -            &        -           &         -          &0.087   $\pm$0.011   &            -      &             -     &           -     &     -          	\\		  	  	  
 \hline
\end{tabular}
\begin{tablenotes}
 \footnotesize
  \item[\emph{a}]{Internal stellar color excess for each source, estimated according to \cite{2000calzetti}.}
 \end{tablenotes}
\end{threeparttable}
}
%\end{center}

%\end{table}
\end{sidewaystable*}

\subsection{Oxygen abundances and radial distances} \label{radial_distances_analysis}

In order to determine the behavior of the metallicity in the system NGC 4656, from GMOS fluxes we have estimated the oxygen abundances using semi-empirical methods through the N2 and O3N2 calibrators, given that our data does not cover the spectral region of temperature sensitive [O{\sc iii}]$\lambda$4363\AA\hspace{0.7mm} emission line. We used the calibrations proposed by \cite{2013marino}, who provided linear relations to obtain oxygen abundances through the N2 and O3N2 calibrator, with an accuracy of 0.16 and 0.18 dex, respectively. Uncertainties in the oxygen abundances have been calculated by propagating the flux uncertainties and adding in quadrature the 0.16/0.18 dex value associated with the scatter in the calibration. The results are shown in detail for each region in Table \ref{table:abundancias_densidad}.

%\subsubsection{Radial projected distances}\label{radia_distances}

The projected galactocentric distances to each star-forming region in NGC 4656 were estimated following the method of \cite{2008scarano}. Given the high inclination of this system \citep[i=82$^{\circ}$,][]{1983stayton}, distances were not corrected by inclination. In the case of the position angle, we used a value of PA=40$^{\circ}$ (taken from \citealt{2012schech}),  which corresponds to the position angle of the major axis seen on Fig. \ref{rgb}. 

Although a normalization of the galactocentric distance (for example, to the optical radius R$_{25}$) is useful to compare the metal distributions between different galaxies, we adopt an absolute radial scale (dex kpc$^{-1}$) for NGC 4656 since this object presents a very disturbed morphology with tidal characteristics, for which the chosen scale has a greater physical significance in this particular case. The results for the projected distances and their respective uncertainties calculated from the uncertainty in the scale of the distance to the source ($\sim$0.024$\pm$0.005 kpc arcsec$^{-1}$), are listed in the second column of Table \ref{table:abundancias_densidad}.  

\subsection{Electron densities}

The electron densities were calculated from the observed [S{\sc ii}]$\lambda\lambda$6716,6731\AA\hspace{0.7mm} ratio (hereafter RS2) using the task {\sc temden}, in the {\sc stsdas nebular} package from {\sc iraf}. All the tasks in this package implement the five-level atomic model FIVEL, developed by \cite{1987derobertis}. Since we are not able to estimate the electron temperature because of the spectral coverage, we assumed an electronic temperature of T$_{e}$= 10.000 K, which is expected for star-forming regions \citep{1989oster}. The values of the electron densities and their respective uncertainties (given by the standard deviation) are presented in the last column of Table \ref{table:abundancias_densidad}.

\subsection{Star formation rates}

There are different star formation rate (SFR) indicators, one of them is the H$\alpha$ nebular emission which emerges from the recombination of gas that has been ionized by the population of most massive stars (O- and B- stars). This indicator traces star formation over the few last million years.
From our flux calibrated Fabry-Perot data, the SFRs for NGC 4656 and its TDG candidate have been estimated through their H$\alpha$ luminosity, using the expression given in \cite{1998kenni}, where:
\begin{equation}
 SFR_{H\alpha} = 7.94 \times 10^{-42} \times L(H\alpha),
\end{equation}
assuming a continuous star formation process.  It is important to note that this SFR can be underestimated due to the fact that H$\alpha$ is affected by extinction. In our case, H$\alpha$ maps have not been corrected by extinction which is variable along the extension of the systems. The uncertainties in the luminosities have been estimated propagating the errors of the fluxes, and considering an uncertainty in the distance to NGC 4656 of 1.1 Mpc \citep{2014sorce}. For the SFRs, the uncertainties were derived propagating the errors for luminosities. These results are presented in Tables \ref{table:luminosidades} and \ref{table:luminosidades2}.

\section{Kinematics}\label{sec:analysis2}
%\subsection{FP Data cubes}

\subsection{Rotation curves}

%\subsubsection{Classical method}
\label{rcclasic}
The rotation curves (RCs) trough the classical method for different regions of NGC 4656 were computed with the software ADHOCw, developed by \cite{1993boul}. These curves are constructed from the velocity fields obtained from the FP data, by taking into account the points located inside a given angular sector from the major axis. On this way we avoid the inclusion of points near the minor axis which show a large dispersion in the computed RC. To estimate the values of velocity and radius, we considered crowns with a width of 25 pixels, and we obtained a rotational velocity and average radius within each crown. For NGC 4656 we considered an angular sector of $\pm$15$^{\circ}$ from the major axis. %, an expected value for highly inclined galaxies \citep{2002garrido}, as is the case. 
To compute the RCs is necessary to adjust some parameters: coordinates of the rotation center, systemic velocity, position angle of the major axis and inclination  of the disk of the galaxy. Since NGC 4656 does not present a clear nucleus, we have adopted as the rotation center the point that provided us the most symmetric RC. %In other cases, the center of symmetry of the velocity field is chosen, which is associated in most cases with the position of the nucleus. We were adjusting the systemic velocity so that when overlapping the approaching and receding sides, both were in agreement.
The systemic velocity was obtained once the optimal superposition between both sides of the RC (approaching and receding sides) was found. The position angle is given counterclockwise from the North.
Due to the irregular shape of the galaxy and its disturbed velocity field, it is not relevant to compute the kinematical inclination of the disk of NGC 4656 (with respect to the plane of the sky); so we have taken different values from literature \citep{1983stayton,2012schech}. The final parameters used in the RC that best fits to NGC 4656 are shown in the Table \ref{table:parametros_curva}.
%You should write how those values have been computed, from morphological consideration ???? P.Amram
\subsection{Position-Velocity diagrams}

The classical method for fitting RCs (see \S \ref{rcclasic}) works properly for low inclination galaxies, however for highly inclined galaxies, as the case of NGC 4656, obtaining RCs through this method is not reliable, since it can lead to large underestimates of the rotation velocities. This is mainly due to the fact that the shape of a profile is obtained from the integration along the line of sight of a large portion of the disk, for which it is necessary to make some corrections to the estimated velocities \citep{2013rosado}. For this reason we have built position-velocity (PV) diagrams which allow obtaining a more reliable RC and measurement of the rotation velocity to systems with high inclination.

In this work, the creation of the PV diagrams was done using the code \textsc{fluxer}\footnote{Package for data cubes in \textsc{idl}, written by Christof Iserlohe. \url{http://www.ciserlohe.de/fluxer/fluxer.html}}. For NGC 4656 we made a mosaic with the 6 data cubes wavelength calibrated, finally forming a single large cube for the whole galaxy (with a total FoV of 5.8'$\times$5.8'). Subsequently, the spectral extraction of the data cube was performed from a slit aligned along the major axis of the system, considering a PA of 40$^{\circ}$, and a slitwidth of 2.5 arcsec. Finally, the PV diagram was built directly from graphical interface of \textsc{fluxer}, which is shown in Fig. \ref{pvs}. The construction of the RC was done from an envelope of the PV diagram, as is explained in detail in \S \ref{sec:resul_pv}.

\subsection{Velocity field modeling}

Considering the complex nature of NGC 4656, we have performed a simple phenomenological model of circular orbits in a plane, following the recipes given by \citet{1991bertola}, we consider the following parametrical axisymmetric rotation curve  

\begin{equation}
{\rm V_{rot}(r)= \frac{V_0\ r}{(r^2+c_0^2)^{\frac{p}{2}}}}
\label{aaaa}
\end{equation}
where $r$ is the radius on galaxy plane; $V_0$ is related to the amplitude of the rotation curve; $c_{0}$ is a concentration parameter. The parameter $p$ defines the form of the rotation curve, varying in the range between 1 (logarithm potential) and 1.5 (Plummer potential, which is a Keplerian-like potential), which is the expected range of values for galaxies. Using the relations linking the velocities in the sky and galaxy planes \citep{1973warner,1978vanderkruit}, the line-of-sight velocity model on sky plane, ${\rm V_{mod}(R,\psi)}$, can be written as: 

\begin{equation}
\resizebox{\columnwidth}{!} 
{$
{\rm V_{mod}(R,\psi)= V_{s}+\frac{V_0\,R\,cos(\psi-\psi_{0})\,sin(i)\,cos^p (i)}{\{R^2[sin^2(\psi-\psi_{0})+cos^2(i) cos^2(\psi-\psi_{0})]+c_{0}^2 cos^2(i)\}^{\frac{p}{2}}}},
$}
\label{eqbertola}
\end{equation}
where $R$ and $\psi$ are the radial and angular projected coordinates 
on sky plane;  V$_{s}$ is the 
systemic velocity; $\psi_{0}$ is the position angle of the line of 
nodes; { $i$} is the disk inclination ($i$ = 0 for face-on 
Tullydisks). Furthermore,  there are two implicit parameters: the coordinates of the kinematic center, $R_{cx}$ and $R_{cy}$. The parameters  $V_0$ and $i$ are degenerates, and hence the values derived, from the fitting, for them are not reliable. Nevertheless, assuming that the ellipticity of the outermost isophote is due to disk inclination we can fix $i$ breaking up the degeneracy. The value for this parameter was taken from \citet{1983stayton} who performed a photometry analysis on NGC4656, they found $i$=82$\degr$ for the galactic disk. The uncertainties associated to barycenter velocity, $\sigma_{V_b}$, can be easily derived by doing an error propagation of the barycenter equation, so we found:

%NOTE THE FOLLOWING YOU GIVE DEPENDS ON THE PARAMETRIC FORM GIVEN ABOVE, SO YOU SHOULD GIVE IT OTHERWISE THE NEXT FORMULA DOES MEAN ANYTHING THIS IS WHY I ADD IT\\

\begin{equation}
{\rm\sigma_{V_b} = \frac{c}{\lambda_{b}}\sigma_{\lambda_{b}}=\sqrt{\frac{\sum_{\lambda_{i}=\lambda_{0}-3*\sigma}^{\lambda_{i}=\lambda_{0}+3*\sigma} I_{i}^2}{(\sum_{\lambda_{i}=\lambda_{0}-3*\sigma}^{\lambda_{i}=\lambda_{0}+3*\sigma}I_{i})^2}}\sigma_{\lambda}}
\end{equation}
 where $c$ is light speed; $\lambda_b$, the measured barycenter wavelength; $\sigma_{\lambda_{b}}$, the uncertainty of the barycenter wavelength; $I_{i}$, the intensity of pixel i; $\sigma$, the measured line dispersion;  and $\sigma_{\lambda} \simeq$ 27 km\,s$^{-1}$ the spectral dispersion for a resolution R$\sim$11000 at H$\alpha$. The mean uncertainty overall FoV is $\sigma_{V_b}=9\pm 3$\,km\,s$^{-1}$. 
We performed the fitting by applying the Levenberg-Marquardt method and taking into account the velocity uncertainties. The parameters fitted of the kinematic model are listed in Table \ref{table_model}.%the xxx, $I_{i}$ the intensity as a function of the wavelength  FOR EACH PIXEL i ??? DOES $\sigma$ DEPENDS ON I??? ; $\sigma$ the measured line dispersion;  and $\sigma_{\lambda} \simeq$ xxxx km\,s$^{-1}$ the spectral dispersion for a resolution R$\sim$11000 at H$\alpha$

%%%%%%%%%%%%%%%%%%%%%%%%%%%%%RESULTS%%%%%%%%%%%%%%%%%%%%%%%%%%%%%%
%%%%%%%%%%%%%%%%%%%%%%%%%%%%%%%%%%%%%%%%%%%%%%%%%%%%%%%%%%%%%%%%%5

\section{Results}  \label{sec:results}

\subsection{Ionizing mechanisms of regions}

\begin{figure}
  \hspace{-0.7cm}
\includegraphics[scale=0.54]{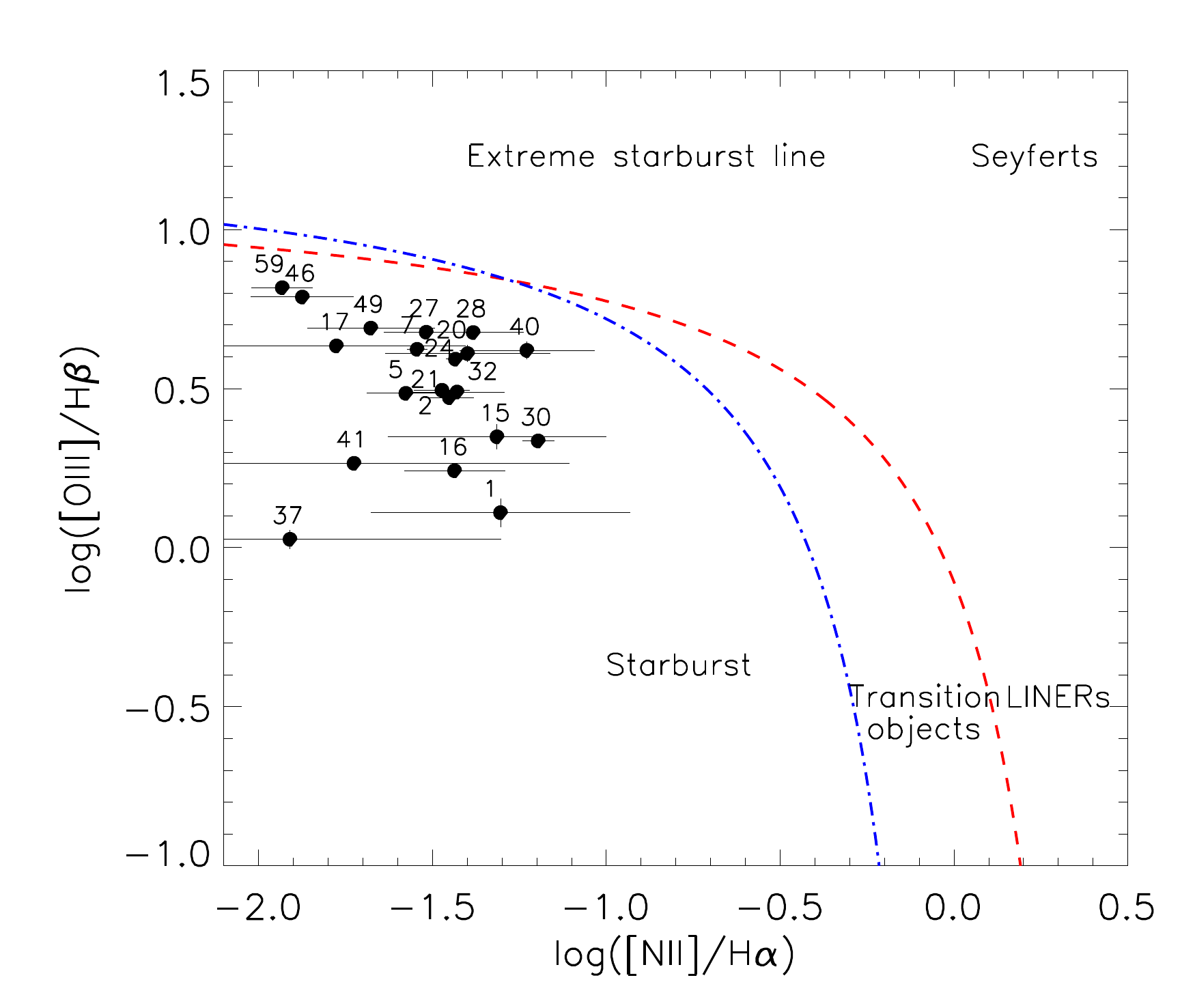}
   \caption{Ionization mechanism diagnostic diagram \citep{1981baldwin} for the regions located in NGC 4656. The blue and red dashed line corresponds to the limit between H{\sc ii} regions and AGNs, as described in \protect\cite{2003kauffmann} and \protect\cite{2001kewley}, respectively. }
   \label{bpt_diagram}
    \end{figure}

Emission-lines intensity ratios are a useful tool for classifying the spectra of extragalactic sources.  
We have constructed the BPT diagnostic diagram \citep{1981baldwin} to confirm the excitation mechanism of the regions observed in NGC 4656, which is presented in Fig. \ref{bpt_diagram}. This plot shows the [O{\sc iii}]/H$\beta$ versus [N{\sc ii}]/H$\alpha$ line ratios, where blue dotted-dashed and red dashed lines correspond to the limit between star-forming objects and active galactic nuclei (AGN), as described by \cite{2003kauffmann} and \cite{2001kewley}, respectively.   

It was not possible to include any region of NGC 4656UV because none presented the four emission lines needed to be placed on the diagram.
On the other hand, there are also some regions of NGC 4656 that are not in the diagram because their emission lines H$\beta$ and/or [N{\sc ii}]$\lambda$6584 were noisy and difficult to measure (regions  $\#$10, 11, 12, 13, 19, 36, 44, 45, 50, 56, 57, 58 and 60). All the regions that we could locate in the BPT diagram (20 in NGC 4656), are located in the area defined by objects dominated by star formation. Therefore, from the BPT diagram, and taking into account the uncertainties in the fluxes, we conclude that the regions to be analyzed correspond to star-forming regions, excluding the AGN activity or ionization by shocks.

\subsection{NGC 4656UV}
\subsubsection{H$\alpha$ emission}\label{sec:tdg_emission}
\begin{figure}
 \hspace{-0.3cm}
\includegraphics[scale=0.85]{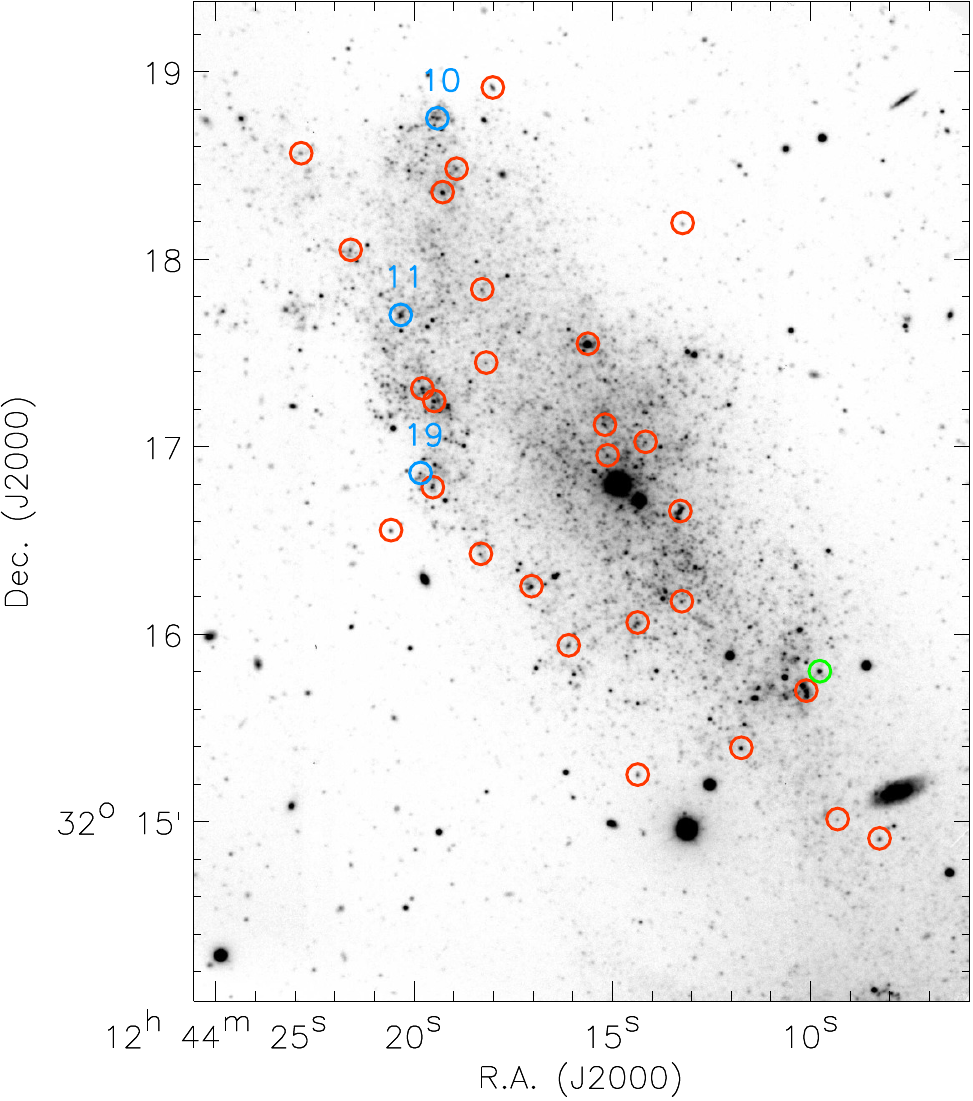}
%  \captionsetup{font=tiny}
   \caption{Gemini/GMOS image in g' filter for NGC 4656UV. Red and blue circles represent regions observed. Blue circles represent the regions which have nebular emission lines, and the red ones the regions that do not contain H$\alpha$ emission. Green circle represents a region with SDSS spectrum. } 
   \label{tdg_regions}
 \end{figure}

\begin{figure*}
 \centering
\includegraphics[scale=0.98]{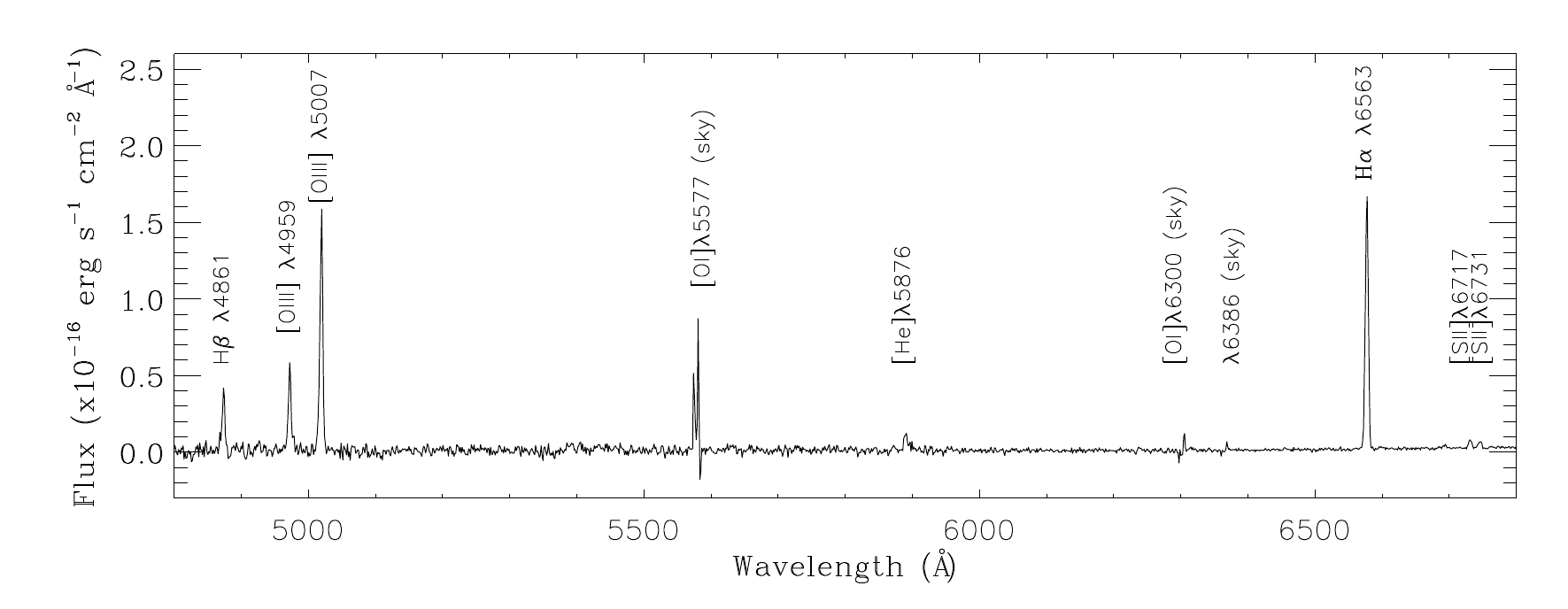}
%\vspace{-0.9cm}
%\includegraphics[scale=0.65]{correspectro11paper.pdf}\\
%\vspace{-0.9cm}
%\includegraphics[scale=0.65]{correspectro23paper.pdf}\\ 
%\caption{Spectra of regions located in NGC 4656UV, where the main emission lines found are labeled. Top panel: Spectrum of the region $\#$19, flux calibrated and corrected by extinction. Intermediate panel: Region spectrum $\#$11, flux calibrated. Bottom panel: Region spectrum $\#$23, flux calibrated.}
\caption{Spectrum of region $\#$19 located in NGC 4656UV, flux calibrated and corrected by extinction. The main emission lines found are labeled.}
\label{espectros_tdg}
    \end{figure*}

Figure \ref{tdg_regions} shows a GMOS image in g'-band for NGC 4656UV. Circles represent the sources observed with GMOS, blue circles indicate the regions with H$\alpha$ emission lines at the redshift of NGC 4656UV, and red circles the rest of them that does not present any detected emission line. The green circle represents a region observed with the SDSS survey. It is important to note that from the 30 regions observed in NGC 4656UV, there are just three of them which exhibit H$\alpha$ emission (regions $\#$10, $\#$11 and $\#$19). From these three regions only source $\#$19, whose spectrum is presented in Fig. \ref{espectros_tdg}, shows the most of emission lines that characterize a H{\sc ii} region (H$\beta$, [O{\sc iii}]$\lambda\lambda$4959,5007\AA,  H$\alpha$ and [S{\sc ii}]$\lambda\lambda$6716,6731\AA), but the [N{\sc ii}]$\lambda\lambda$6548,6584\AA\hspace{0.7mm} lines were not detected despite the fact that with a resolution R=875,
they should be resolved and distinguishable from H$\alpha$.% (by the way how broad is Halpha if you measure it at the continuum level, just to be sure there is no pb with the resolution?) by P.Amram
 %In Fig. \ref{espectros_tdg}, the top panel shows the spectrum of region $\#$19 which has several emission lines, the middle panel shows the spectrum of region $\#$11 which presents H$\alpha$ emission and is very similar to spectrum $\#$10, and the lower panel shows the spectrum of region $\#$23, which represents the typical type of spectrum found in the rest of NGC 4656UV. This last spectrum does not present strong nebular emission lines and is dominated by the noise, as it is also the case for the 26 remaining spectra. For this reason, in this work it was not possible to determine the physical properties of NGC 4656UV resolutely.%, so for this reason in this work it was not possible to determine the physical properties of the TDG candidate (electronic density, oxygen abundance gradient, age of regions, etc.). esto es muy negativo y mas para la tesis 

\begin{figure}
 \centering
\includegraphics[scale=0.75]{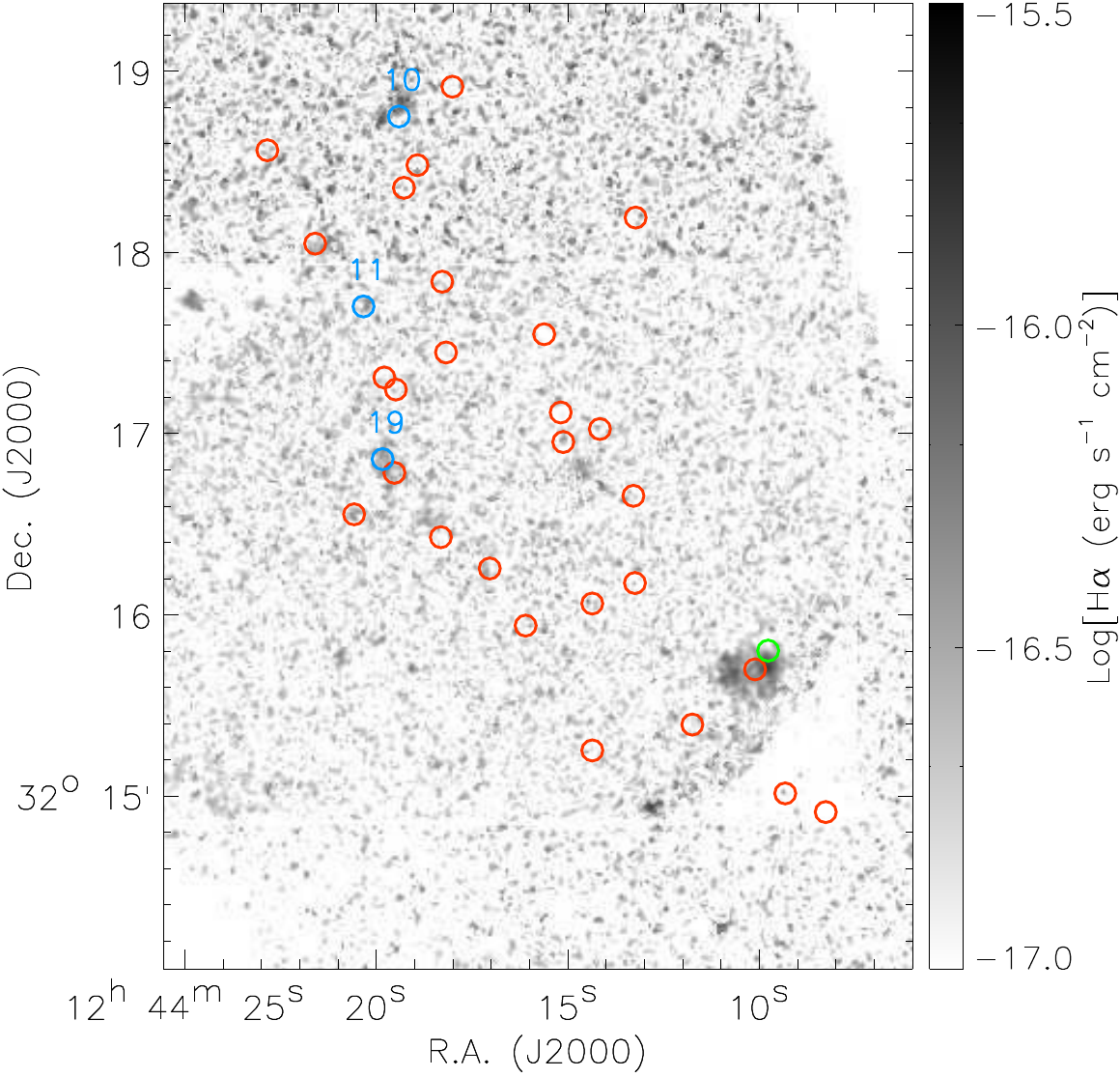}
   \caption{Monochromatic H$\alpha$ emission map for NGC 4656UV. The circles represent GMOS regions under the same color code of Fig. \ref{tdg_regions}. The H$\alpha$ intensity is shown in logarithmic scale.}
   \label{tdg_mono} 
\end{figure}

The weak H$\alpha$ emission of the TDG candidate is confirmed by the monochromatic H$\alpha$ map obtained from FP data, as shown in Fig. \ref{tdg_mono} where circles represent regions observed with GMOS under the same color code than Fig. \ref{tdg_regions}. We can clearly see that in the location where the TDG candidate is supposed to be, there is practically no H$\alpha$ emission ($<$10$\times$10$^{-16}$ erg s$^{-1}$cm$^{-1}$). The maximum emission is observed beside the region observed by SDSS (green circle), its  flux is about 10$^{-16}$ erg s$^{-1}$cm$^{-2}$ per pixel, which is an order of magnitude higher than in the rest of the system ($\sim$ 10$^{-17}$ erg s$^{-1}$cm$^{-2}$ per pixel). Due to this weak emission in NGC 4656UV, it was not possible to study its kinematics, however it was possible to estimate its oxygen abundance (see \S \ref{sec:tdg_abundance}) and SFR (see \S \ref{sec_sfr}). 

\subsubsection{Oxygen abundance}\label{sec:tdg_abundance}

\begin{figure*}
 \centering
  \includegraphics[scale=0.98]{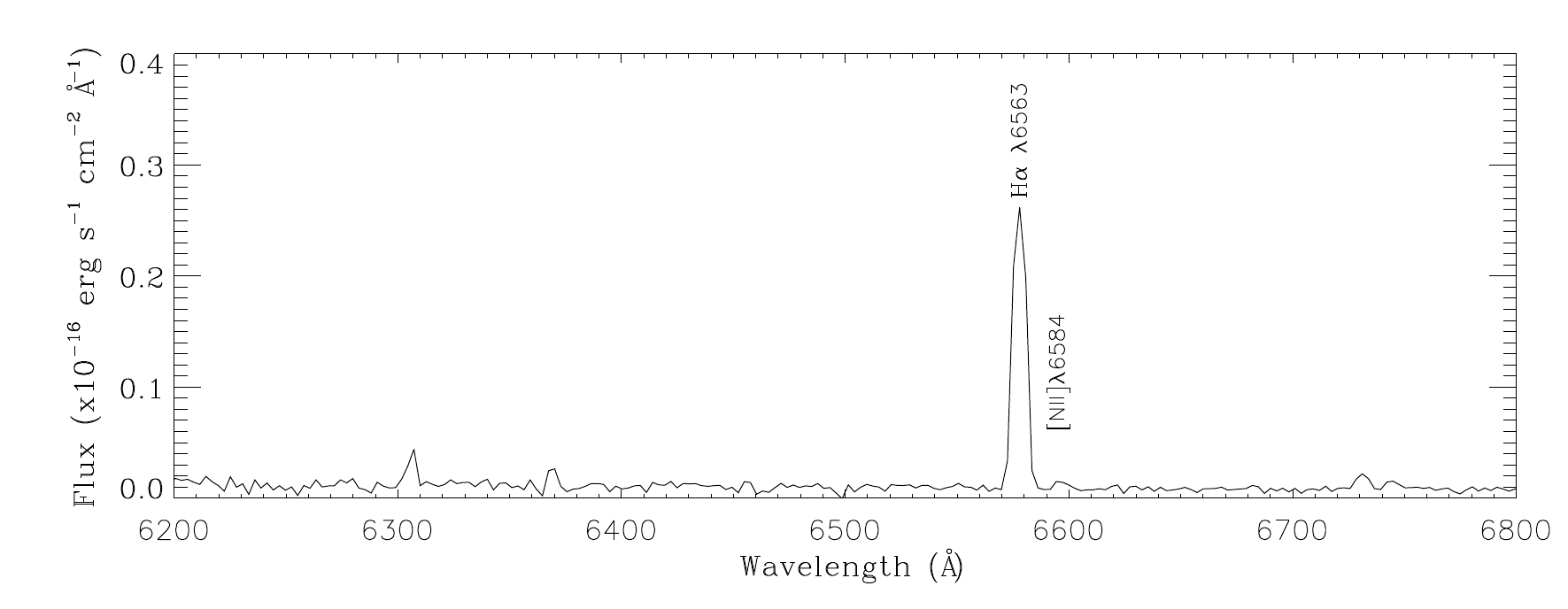}
 \caption{Flux calibrated spectrum of region $\#$11 located in NGC 4656UV, showing the H$\alpha$ emission line.}
% \includegraphics[scale=0.75]{espectro11tdg_binning3.pdf}\\
 %\vspace{-1.1cm}
%\includegraphics[scale=0.75]{correspectro11_zoom.pdf}
%\caption{Spectra of region $\#$11 located in NGC 4656UV. Top panel: Zoom of the flux calibrated spectrum after going through a procedure of \textit{re-binning}. It is possible to appreciate the presence of the [N{\sc ii}]$\lambda$6584 emission line. Bottom panel: Zoom of the flux calibrated spectrum.}
\label{espectro_tdgbin}
    \end{figure*}

%As described in \S \ref{sec:tdg_emission}, after the GMOS spectra extraction process for NGC 4656UV, no region with emission was obtained in both lines H$\alpha$ and [N{\sc ii}]$\lambda$6584$\AA$ at full resolution, which allows an estimation of the  oxygen abundance through the N2 calibrator.
%However, it was possible to find the emission line [N{\sc ii}]$\lambda$6584$\AA$ in the spectrum of region $\#$11 after a re-binning procedure. We resized the bin between each element of the wavelength axis, decreasing it to half its original size through neighborhood averaging. The resulting spectrum is in the top panel of Fig. \ref{espectro_tdgbin}. As a comparison, in the lower panel of Fig \ref{espectro_tdgbin} the spectrum of region $\#$11 is shown before the re-binning process. 
We have estimated the oxygen abundance for NGC 4656UV from region $\#$11 once its spectrum was binned. We resized the bin between each element of the wavelength axis, decreasing it to half its original size through neighborhood averaging. The resulting spectrum is shown in Fig. \ref{espectro_tdgbin}. We have obtained an oxygen abundance of  12+log(O/H)$\sim$8.03$\pm$0.20, through the N2 calibrator \citep{2013marino}. With the same calibrator, we have also calculated the oxygen abundance for another region of NGC 4656UV observed by SDDS (Fig. \ref{tdg_regions}, green circle), considering the fluxes tabulated in the SDSS DR12 database, and we obtained a value of  12+log(O/H)$\sim$8.34$\pm$0.20. Both oxygen abundances corresponding to different zones of the system are similar, considering the uncertain in the fluxes and in the \cite{2013marino} calibrations.

\subsection{NGC 4656}
%\subsection{H alpha emission}
\subsubsection{H$\alpha$ emission}
%AGREGAR FIGURA GMOS CON TODAS LAS REGIONES?

\begin{figure}
   \centering
\includegraphics[scale=0.53]{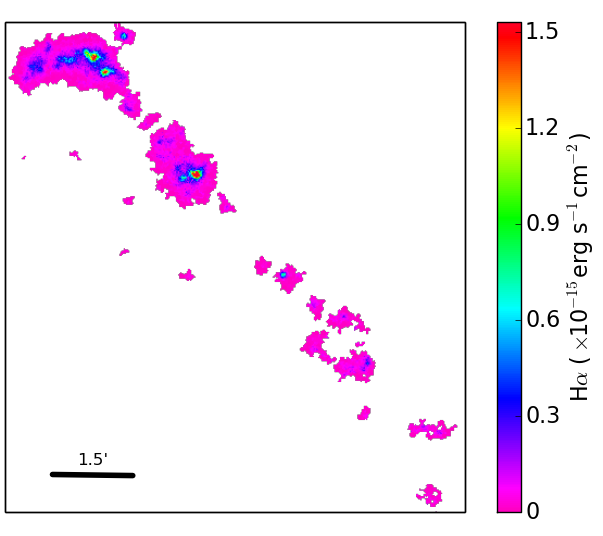} 
   \caption{H$\alpha$ monochromatic map for NGC 4656. The color bar represents the flux in physical units.}
   \label{ngc4656_mono}
    \end{figure}

The H$\alpha$ monochromatic map for NGC 4656 is shown in Fig. \ref{ngc4656_mono}. The most intense H$\alpha$ emission is concentrated in the north-east part of the galaxy, region which is dominated by H{\sc ii} regions  surrounded by diffuse emission. These H{\sc ii} regions have a size of the order of hundreds of parsecs ($\sim$150-650 pc). As we move towards the southwest, the number of H{\sc ii} regions decreases and diffuse emission predominates. These results are consistent with what is shown by the GMOS spectra.%At the northeastern end of the disk they make a hook-like appendix, extending by more than 1 to the east. The SW part of the disk shows very weak Hα emission and very few small H ii regions. We also note a weak ridge of Hα emission running parallel to the disk plane at a distance of about 1' towards the SE. In the NE half of the disk, the space between this ridge and the disk plane is filled with faint diffuse spurs. (paper magnetic fields)
%\subsubsection{H$\alpha$ profiles}

%In Figs. x and x, we display a few examples of H$\alpha$ profiles throughout different areas of the system.\\

\begin{figure*}
\centering
\includegraphics[scale=0.27]{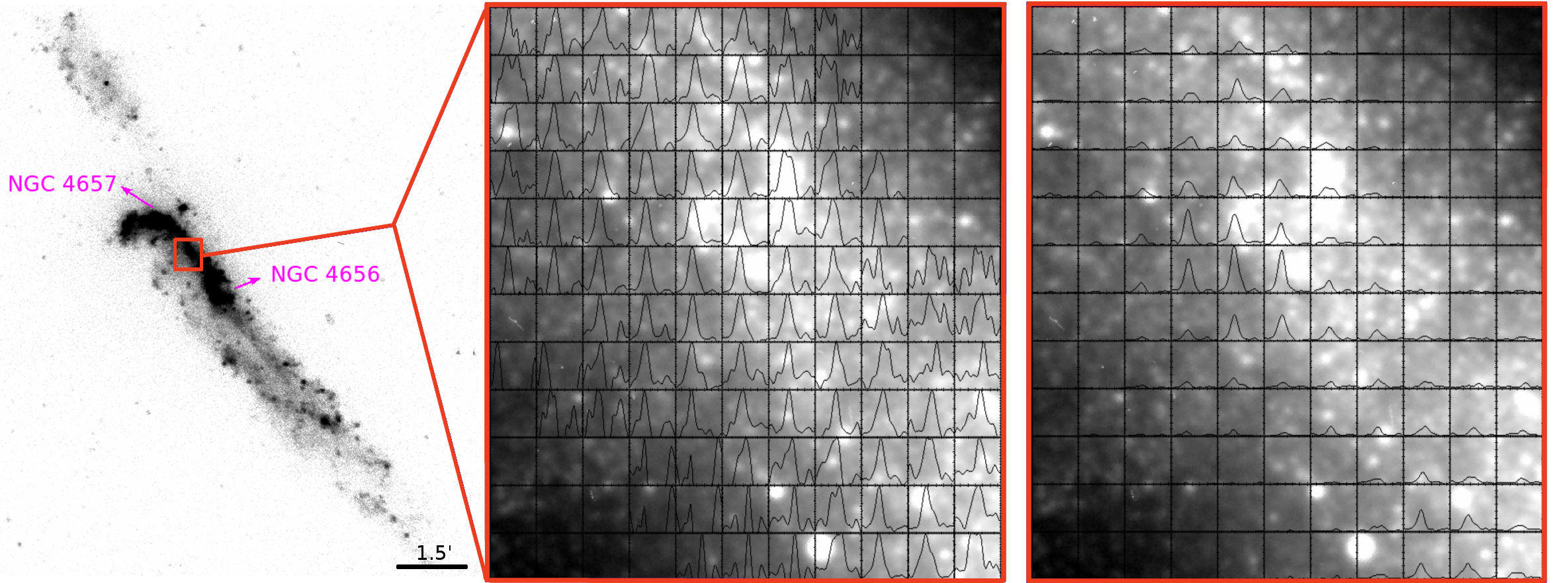}
\caption{Left: Image in the FUV band of NGC 4656/4657, observed with \textit{GALEX}. The regions that have been designated by some authors as NGC 4657  (northeast sector of the system) and NGC 4656 (southwest sector of the system) are indicated with arrows. The red square corresponds to the area whose profiles are shown in the following panels. Center: H$\alpha$ profiles obtained after doing a $binning$ of 6$\times$6 pixels. Each ``each'' that contains each profile  has dimensions of $\sim$4.1''$\times$4.1'' ($\sim$100$\times$100 pc$^{2}$). The size of the total region shown is $\sim$50''$\times$45'' ($\sim$1.2$\times$1.0 kpc$^{2}$). Each profile is normalized to the size of each box. The background image was observed by GMOS in the \textit{g}$'$ band. Right: H$\alpha$ profiles equivalent to those shown in the left panel. The difference is that in this case the profiles are normalized to the brightest profile shown in this area ( $\sim$3.0$\times$10$^{-17}$erg s$^{-1}$cm$^{-2}$). The background image is the same as shown in the left panel.}
\label{perfiles_horizontal}
\end{figure*}

On the northeast edge of NGC 4656, there is a feature which gives the galaxy a ``hockey stick'' appearance \citep{2012schech} also known as NGC 4657.
In order to determine whether the NGC 4656 system is a single disturbed galaxy or two galaxies at some stage of interaction (NGC 4656/4657), we analyzed the H$\alpha$ emission profiles  in the connecting area of the two possible galaxies. This area is highlighted in the left panel of Fig. \ref{perfiles_horizontal}, through a red square. The H$\alpha$ profiles corresponding to this zone are presented in the central and right panels. In the central panel the H$\alpha$ profiles have been normalized to the size of the box and in the right panel the same profiles have been normalized to the brightest profile. The profiles in this region are presented as asymmetrical, wide and some with clear multiple peaks of emission. Wide and asymmetric profiles may correspond to lines formed by several components, which cannot be resolved because the spectral resolution is not high enough (R$\sim$11000). Double profiles can be due to star formation in one single galaxy, for example to the presence of an expanding bubble, or can be due to a merging of two systems, where a profile superposition along the line-of-sight is observed. This kind of analysis has been performed by \cite{2007amram} using Fabry-Perot data, who concluded that the A+C system in HCG 31 is in an early stage of merging, based on the presence of double profiles of the H$\alpha$ line in the region which connects components A and C. However, in our case, the most disturbed profiles that can be seen in the area have a very low emission ($<$1$\times$10$^{-17}$ erg s$^{-1}$ cm$^{-2}$), as presented in the right panel of Fig. \ref{perfiles_horizontal}. Therefore, these profiles do not show clear evidence indicating the overlapping between two galaxies. The symmetric shape of the intense H$\alpha$ profiles in the studied area does not suggest the presence of multiple intense components, which could suggest the presence of overlapping profiles and therefore, galaxies in process of merger.\\ 

\subsubsection{Oxygen abundances and its distribution}

%We have estimated the oxygen abundances for sources observed with GMOS  through the  N2 and O3N2 calibrators, however it was not possible to estimate them in regions which presented N2 and/or O3N2 index out of the valid range\footnote{The linear fit determined by \cite{2013marino} to obtain metallicity of oxygen using the N2 calibrator, with a confidence level of 68$\%$, presents an accuracy of $\pm$0.16 dex over a valid range of -1.6$<$N2$<$-0.2}. The results are shown in detail for each region in Table \ref{table:abundancias_densidad}, with their respective uncertainties which were estimated considering the uncertainties in the fluxes and the dispersion of 0.16 and 0.18 dex for the N2 and O3N2 calibrators, respectively \citep[see][]{2013marino}. 

Oxygen abundance estimates are listed in Table \ref{table:abundancias_densidad}. Given the spectral coverage of the observations and the valid range of the calibrators, most of the presented values were obtained through the N2 method. Then, the following analysis concerning oxygen abundances refer to the N2 estimates. We found oxygen abundances ranging between 12+log(O/H)=8.03 (regions $\#$5 and $\#$7, outwards) and 12+log(O/H)=8.34 (region $\#$44, near the center). The average chemical abundance for NGC 4656 is 12+log(O/H)=8.10$\pm$0.20.

\begin{figure*}
   \centering
\includegraphics[scale=0.55]{{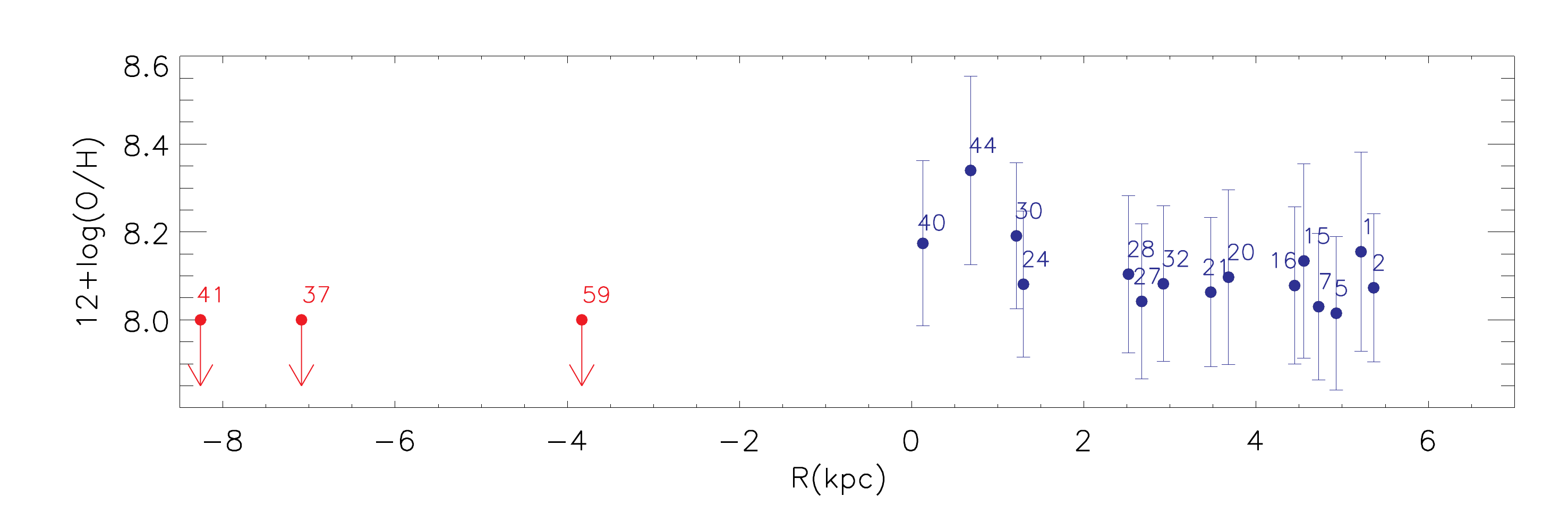}} \\
\vspace{-0.6cm}
\includegraphics[scale=0.55]{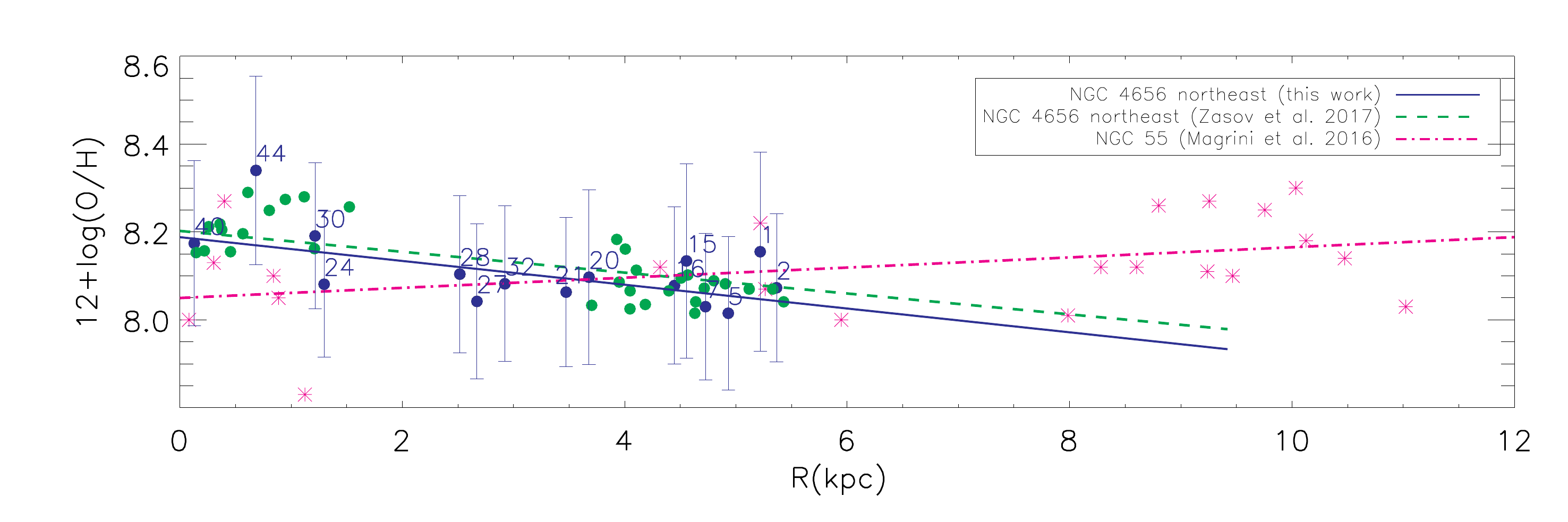}
   \caption{Top panel: Blue dots show the oxygen abundance values obtained for the regions located at the northeast of NGC 4656, with the N2 calibrator \citep{2013marino}. Red points represent the values for regions located at the southwest of NGC 4656 which have N2 index out of range (see footnote in previous page), therefore their metallicity are less than 8.0. The latter is represented by red arrows. Bottom panel: Top panel zoom. Blue line represents the metallicity gradient for NGC 4656 regions obtained in this work. The green dashed line represents the oxygen abundance gradient estimated with the N2 calibrator \citep{2013marino} for the NGC 4656 regions obtained by \protect\cite{2017zasov} (green circles). For comparison, the dotted-dashed magenta line represents the oxygen abundance gradient estimated with the N2 calibrator  \citep{2013marino} for the NGC 55 regions obtained by \protect\cite{2016magrini} (asterisks magenta).}
   \label{abundancias}
    \end{figure*}

%\subsubsection{The oxygen abundance gradient in NGC 4656}
Fig. \ref{abundancias} shows the oxygen abundance distribution in NGC 4656. On the top panel,  blue dots represent the oxygen abundances obtained for regions located at the northeast region of the galactic center, with their respective identification number (see Fig \ref{rgb}). The red dots represent the sources observed at the southwest region of NGC 4656. All these sources have N2 indexes out of the valid range of this calibrator. It suggests that these sources have  abundances lower than 12+log(O/H)$\sim$8, which is indicated with red arrows in Fig. \ref{abundancias}. From the top panel of Fig. \ref{abundancias}, we can note that the metal distribution along NGC 4656 is not symmetric with respect to its center, with oxygen abundances lower than 12+log(O/H)$\sim$8 in the southwest sector, and greater than this value in the northeast sector.

% By way of comparison, we also show the results obtained by \cite{2008pilyugin} for three regions located in NGC 4656, represented by orange points. It is necessary to point out that these authors calculated the oxygen abundance through a different method from the one used in this work, which is based on the auroral lines [O{\sc ii}]$\lambda$732 and 733 nm (modification of the standard method T$_ {e}$). 
The bottom panel of Fig. \ref{abundancias} corresponds to the oxygen abundance distribution obtained for the northeast part of NGC 4656 (blue points in the upper panel), where the solid blue line represents a linear fit applied to the data which has been obtained using the \textsc{idl} routine \textsc{mpfitexy} \citep{2010williams}, which considers the uncertainties in both axis. As a comparison, on this plot, we have included the oxygen abundances estimates (N2 method) that we have derived from the emission line fluxes published by \cite{2017zasov} for this system, and whose distances were re-calculated by us by following the method described in \S \ref{radial_distances_analysis} (we note that \citealt{2017zasov} published oxygen abundances derived from IZI and S methods). These values are represented by green points and the green dashed line shown a linear fit on the data. In addition, on this plot we have included the oxygen distribution for NGC 55, which is a nearby galaxy that displays tidal features and similar characteristics than NGC 4656 (NGC 55  is located at 2.34 Mpc, with a total mass of 2.0$\times$10$^{10}$M$_{\odot}$ and R$_{25}$=11 kpc, \citealt{2013west,2016kudri}). In order to analyze its oxygen distribution, we used the abundances that were recently published by \cite{2016magrini}, who used the N2 calibrator proposed by Marino et al. (2013). Finally, projected distances for this system were re-calculated by using the positions given by \cite{2016magrini} and the method described in \S \ref{radial_distances_analysis}. Then, the gradient for NGC 55 is represented by magenta asterisks and a dashed-dotted magenta line. 

%According to the linear fit applied to oxygen abundances (see lower panel Fig. \ref{abundancias}), 

In the case of our measurements for NGC 4656, we found a zeropoint of 12+log(O/H)=8.19$\pm$0.12 and a slope of $\beta$=-0.027$\pm$0.029 dex kpc $^{-1}$ for the northeast sector of the galaxy, indicating an approximately homogeneous oxygen abundance distribution compared to isolated spiral galaxies, which could display a typical gradient of $\sim$0.072 dex kpc $^{-1}$ \citep{1994zari}. It should be noted that the value of $\beta$ found for NGC 4656 must be taken with caution due to the tidal morphology that it presents. In addition, we found a good agreement between our estimates and the oxygen distribution derived from the fluxes measured by \cite{2017zasov} (average 12+log(O/H)$\sim$8.13, slope -0.023 dex kpc$^{-1}$). This agreement even exists when we compared our measurements (N2 method) with respect to the values obtained by \cite{2017zasov} through the theoretical IZI method (average 12+log(O/H)$\sim$8.18, $\beta$=-0.03 dex kpc$^{-1}$). The oxygen distribution obtained in this work also agrees with the gradient obtained by \cite{2014pilyugin} (gradient --0.03 dex kpc$^{-1}$) following the C method \citep{2012pilyugin}. 
%On the other hand, in this work we found that oxygen abundance values are lower in the northeast sector of NGC 4656 than in its optical center, according to \cite{2017zasov}, however this difference is within the uncertainty.

On the other hand, unlike the gradient reported in this paper for NGC 4656, the radial metallicity gradient for NGC 55 has a positive slope (+0.011 dex kpc $^{-1}$). 
Despite this difference, both galaxies show flat oxygen abundance gradients considering the uncertainties. We can note, furthermore, that both NGC 4656 and NGC 55 reach large radial distances ($\gtrsim$6 kpc and $\gtrsim$11 kpc, respectively) considering their low masses, being that in general irregular dwarf galaxies (10$^{8}$-10$^{10}$M$_{\odot}$) have diameters between 1 and 10 kpc \citep{2012matteu}. For the case of NGC 4656, the furthest galactocentric projected distance calculated in this work is $\sim$7 kpc (see Table \ref{table:flujos_posiciones}), which is very interesting considering that its effective radius is only $\sim$1 kpc \citep{1973nilson}. This fact indicates that the gas flows can produce a mixture of metals and the redistribution of them reaching large distances \citep{2015bresolin}, showing how powerful the effect of an interaction in the gas mixture can be. %On the other hand, in \S \ref{sec:curves} we have observed, from the velocity field of the galaxy, the existence of kinematic perturbations through the presence of non-circular motions, which can be linked to gas flows.

\begin{table*}
 \centering
 \resizebox{0.7\textwidth}{!}{
 \begin{threeparttable}
 \caption{Distances, oxygen abundances and electron densities for H{\sc ii} located in NGC 4656}\label{table:abundancias_densidad}
 %\vspace{-0.7cm}
 \begin{tabular}{lccccccc}
 \hline\hline  
 ID & Distance\tnote{\emph{a}}& N2 & O3N2 & 12 + log(O/H)\tnote{\emph{b}} & 12 + log(O/H)\tnote{\emph{c}} & RS2\tnote{\emph{d}}&  N$_{e}$ \\
 &Kpc & & & &  & &cm$^{-3}$\\ \hline \\
 1 &5.22$\pm$1.15 &-1.27$\pm$0.33 & 1.39$\pm$0.34 & 8.15$\pm$0.23 & 8.24$\pm$0.20 & 1.33 &94 \\ 
 2 &5.36$\pm$1.18 &-1.45$\pm$0.07 & 1.92$\pm$0.07 & 8.07$\pm$0.17 & -  &1.37 &54.37 \\ 
 5 &4.93$\pm$1.09 &-1.58$\pm$0.11 & 2.06$\pm$0.11 & 8.01$\pm$0.18 & - &1.31 &110.09 \\ 
 7 &4.73$\pm$1.04 &-1.54$\pm$0.03 & 2.17$\pm$0.03 & 8.03$\pm$0.17 & -  &1.41&16.42 \\ 
 %12 & - & - & -& - & - \\ 
 %13 & - & - & -& -  & - \\ 
 15 &4.55$\pm$ 1.01&-1.32$\pm$0.32 & 1.66$\pm$0.32 & 8.13$\pm$0.22 & 8.18$\pm$0.19  & 1.33 &90 \\ 
 16 &4.45$\pm$0.98 &-1.44$\pm$0.14 & 1.67$\pm$0.15 & 8.08$\pm$0.18 & 8.17$\pm$0.18  & 1.38 &40\\ 
 17 &3.85$\pm$0.85 &-1.77$\pm$0.37 & 2.40$\pm$0.37 &-  & -& 1.59 & $<$10 \\ 
 20 &3.68$\pm$0.81 &-1.39$\pm$0.24 & 2.01$\pm$0.24 & 8.09$\pm$0.20 & - & 1.17 &275\\ 
 21 &3.47$\pm$0.77 &-1.47$\pm$0.08 & 1.97$\pm$0.08 &8.06$\pm$0.17 & -  &1.41 &19\\ 
 24 &1.30$\pm$0.29 &-1.43$\pm$0.03 & 2.03$\pm$0.03 & 8.08$\pm$0.17 & -  & 1.26 &160\\ 
 27 &2.67$\pm$0.59 &-1.52$\pm$0.12 & 2.19$\pm$0.12 & 8.04$\pm$0.18  & -& 1.15 &304\\ 
 28 &2.52$\pm$0.56 &-1.38$\pm$0.14 & 2.06$\pm$0.15 & 8.10$\pm$0.18 & - & 1.43& $<$10\\ 
 30 &1.21$\pm$0.27 &-1.19$\pm$0.05 & 1.53$\pm$0.05 & 8.19$\pm$0.17 & 8.21$\pm$0.18 & 1.45 &$<$10\\ 
 32 &2.92$\pm$0.65 &-1.43$\pm$0.14 & 1.92$\pm$0.14 & 8.08$\pm$0.18  & -  & 1.32 &98 \\ 
 40 &0.13$\pm$0.03 &-1.23$\pm$0.19 & 1.85$\pm$0.20 & 8.17$\pm$0.19 & -  &1.39 &37 \\ 
 44 &0.68$\pm$0.15 &-0.86$\pm$0.10 & -             & 8.34$\pm$0.21 & -  &1.16 &303 \\ 
 46 &0.44$\pm$0.10 &-1.87$\pm$0.15 & 2.66$\pm$0.15 & - & -  &1.28 &141\\ 
 49 &0.24$\pm$0.05 &-1.67$\pm$0.18 & 2.37$\pm$0.18 & - & - &1.53 & $<$10\\ 
 %16 & - & - & - & -  & - \\ 
 %17 & - & - & - & -  & -\\ 
 36 &5.78$\pm$1.28 &-              & - & - & -  & 1.57&$<$10\\ 
 37 &7.09$\pm$1.56 &-1.92$\pm$0.62 & 1.95$\pm$0.62 & - &   &1.97 &$<$10\\ 
 41 &8.26$\pm$1.82 &-1.74$\pm$0.64 & 2.00$\pm$0.64 & - & - & 1.25&174\\ 
 %45 & - & - & - & - & -\\ 
 %46 & - & - & - & - & -\\ 
 %50 & - & - & - & -& -\\ 
 59 &3.83$\pm$0.85 &-1.93$\pm$0.09 & 2.75$\pm$0.09 & - & - &1.44 &$<$10\\ \hline
 %60 & - & - & - & -  & -\\ \hline
 \end{tabular}
 \begin{tablenotes}
 \footnotesize
  \item[\emph{a}]{Projected galactocentric distance, corrected according to the method of \cite{2008scarano}.}
 \item[\emph{b}]{Oxygen abundances estimated using the N2 calibrator proposed by \cite{2013marino}, which is valid in the range -1.6$<$N2$<$-0.2.}
 \item[\emph{c}]{Oxygen abundances estimated using the O3N2 calibrator proposed by  \cite{2013marino}, which is valid in the range -1.1 <O3N2 <1.7.}
 \item[\emph{d}]{Ratio between emission lines [S{\sc ii}]$\lambda$6717\AA\, and [S{\sc ii}]$\lambda$6731\AA}
 \end{tablenotes}
 \end{threeparttable} 
 }
\end{table*}

\subsubsection{Electron densities}

Emission line ratios (RS2) and electron densities for the H{\sc ii} regions located in NGC 4656, are presented in the penultimate and last column of Table \ref{table:abundancias_densidad}, respectively.

We found that 15 sources have a RS2 index that is lower than the theoretical limit (RS2=1.43 for a T$_{e}$ = 10000 K, \citealt{1989oster}). For these sources it was possible to estimate the electron densities, which range between 16<N$_{e}$<304 cm$^{-3}$, with an average of N$_{e}$=128. On the other hand, 7 regions present a RS2 value greater than 1.43. In this case, it can be assumed that N$_{e}$ is lower than 10 cm$^{-3}$, however the estimation of the electron density becomes uncertain \citep{2013lopez}. Several authors have faced this same situation, using different kind of data \citep{1989kenni,1994zari,2005bresolin,2013lopez}. For example, \cite{1989kenni}, through long slit spectra, found that numerous H{\sc ii} regions  in spiral galaxies have a RS2 greater than the theoretical limit. On the other hand, using integral field spectroscopy (IFS), \cite{2013lopez} found a similar behavior for sulfur lines in a sample of H{\sc ii} regions in M33.

In summary, we find that the H{\sc ii} regions belonging to NGC 4656 present electron densities of the order of N$_{e}$$\sim$10-300 cm$^{-3}$, which is consistent with the typical range found for other extragalactic H{\sc ii} regions, which have typical densities of N$_{e}$$\leq$500 cm$^{-3}$. Indeed, 13 sources (of a total of 22) present values of N$_{e}$ that are in the range estimated by \cite{2014krabbe} for H{\sc ii} regions located in interacting galaxies (N$_{e}$=24-532 cm$^{-3}$), which is consistent with the stage of the system NGC 4656. 

\subsection{Mass-Metallicity relation}

A useful tool for understanding the formation and evolution of galaxies is the mass-metallicity relation (MZR). Using data from the Sloan Digital Sky Survey (\textit{SDSS}), several authors have confirmed that there is a correlation between the stellar mass of galaxies and metallicity in gas phase \citep{2004tremonti,2008kewley,2014salim}. Interestingly, some authors (e.g. Weilbacher et al. 2003) have shown that TDGs candidates do not follow the MZR relation defined by primordial dwarf galaxies. Therefore, by studying the location of NGC 4656/4656UV in the MZR diagram can given us important clues about the origin of these galaxies. 

In order to examine location of NGC 4656 and NGC 4656UV in the MZR relation, we have considered as a control sample the study carried out by \cite{2015jimmy}, who analyze the low-mass regime of this relation. Given that \cite{2015jimmy} derived oxygen abundances from the N2 calibration proposed by \cite{2002denicolo}, we have recomputed the nebular abundances using this calibrator, yielding 12+log(O/H)=8.07 and 12+log(O/H)=8.26 for NGC 4656 and NGC 4656UV, respectively. 

The stellar mass for NGC 4656 was derived from its near-IR emission (H and K-band, \citealt{2012schech}) and from an averaged M/L ratio derived from nine different colors\footnotetext{The nine colors used were: (g'-r'), (u'-g'), (u'-r'), (u'-i'), (u'-z'), (g'-i'), (g'-z'), (r'-i') y (r'-z'). These were estimated considering the fluxes listed in \citep{2012schech} (Table 2).} (see \citealt{2003bell}). We obtained M/L$_{H}$=0.72, M/L$_{K}$=0.66, which translate into an average stellar mass of M$_{*}$=8.57$\times$10$^{8}$M$_{\odot}$. In the case of NGC 4656UV, the stellar mass was derived by using its g'-band luminosity, given the lack of good photometry in the near IR. In this case we obtain a stellar mass of M$_{*}$=5.46$\times$10$^{7}$M$_{\odot}$.

\begin{figure}
\hspace{-0.5cm}
\includegraphics[scale=0.55]{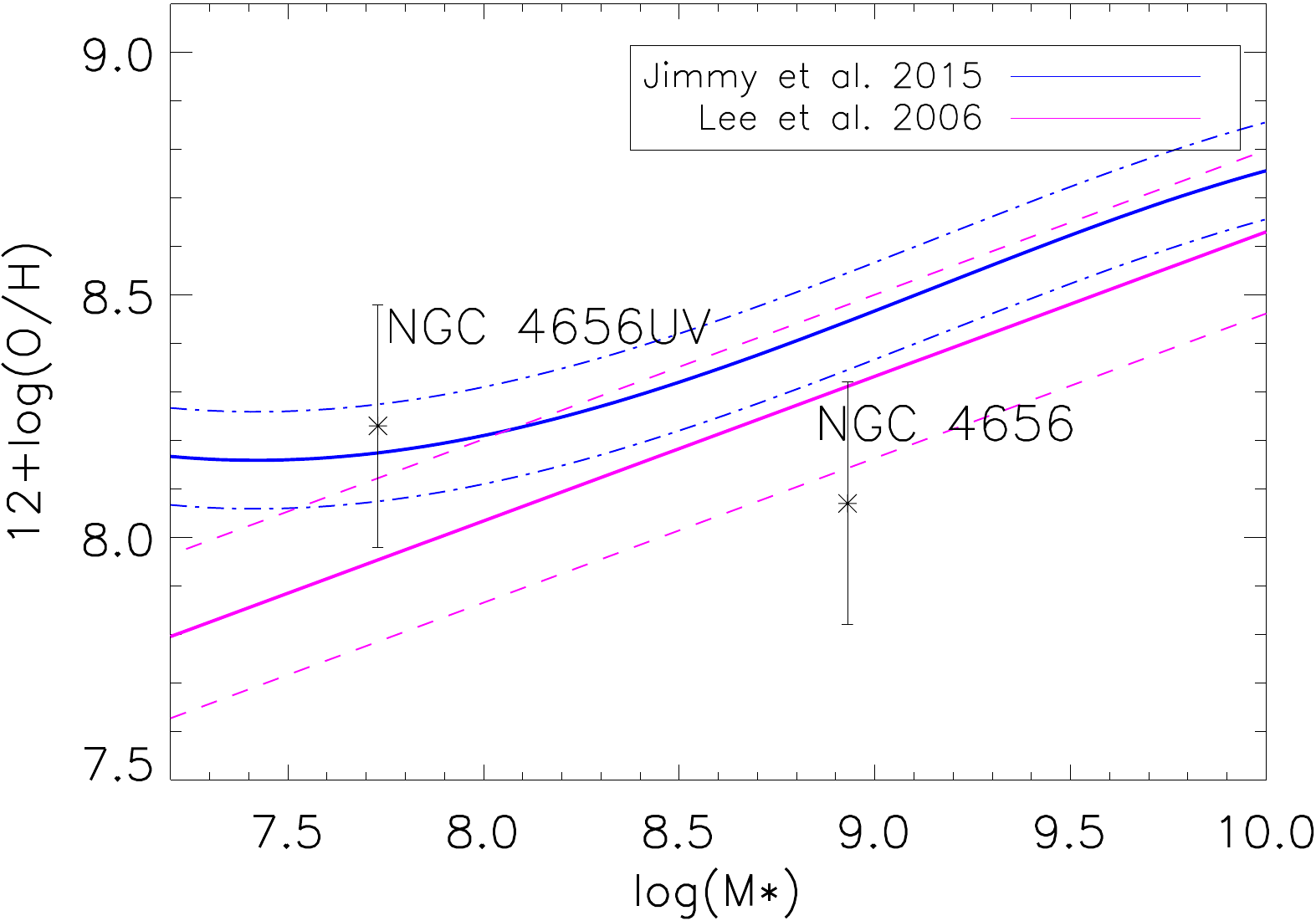} 
\caption{Mass-metallicity relation for NGC 4656 and NGC 4656UV. The blue solid line represents the MZR estimated by \protect\cite{2015jimmy} for normal dwarf galaxies, the dot-dashed line represents the error of the fitting at $1\sigma$. The solid magenta line represents the MZR estimated by \protect\cite{2006lee} for dwarf galaxies with star formation, the dotted line represents the error of the fitting at $1\sigma$.}
\label{MZ}
\end{figure}

In Fig. \ref{MZ} we show the MZR studied by \cite{2015jimmy} (blue lines), where we have included the data for NGC 4656 and NGC 4656UV (black asterisks). The error bars in the chemical abundances represent an uncertainty of 0.25 dex, which arises from the scatter in the calibration (uncertainties updated by \cite{2005perez} based on the calibration proposed by \cite{2002denicolo}). In addition, magenta lines represent the MZR found by \cite{2006lee} for a sample of dwarf irregular galaxies, where abundances were obtained through the direct method. As shown in Fig. \ref{MZ}, NGC 4656UV follows the relation defined by normal dwarf galaxies (\citealt{2015jimmy}), contrary which is expected for TDGs. Indeed, for newly formed TDGs, in which the \textit{in situ} production of metals is insignificant compared to the metals coming from the interacting progenitor galaxies, it is expected that they do not follow a MZR \citep{1998duc}. In this context, this result suggests that the TDG candidate would not have a tidal origin. In the case of NGC 4656, we found that this object lies below the relation proposed by \cite{2015jimmy}, with an abundance very similar to the value found for NGC 4656UV.

%satisfies the relation proposed by \cite{2006lee}, considering the uncertainties. 

\subsection{Star formation rates from H$\alpha$ luminosity}\label{sec_sfr} 

%Using the flux calibrated H$\alpha$ monochromatic maps (Fig. \ref{tdg_mono} and Fig. \ref{ngc4656_mono}), we have estimated the luminosities and  SFRs for NGC 4656 and NGC 4656UV, considering the distance assumed throughout this work \citep[5.1 Mpc,][]{2014sorce}. 

% LO ANTERIOR YA LO HABIAS DICHO EN EL ANALSIS.

\begin{figure}
\centering
\includegraphics[scale=0.6]{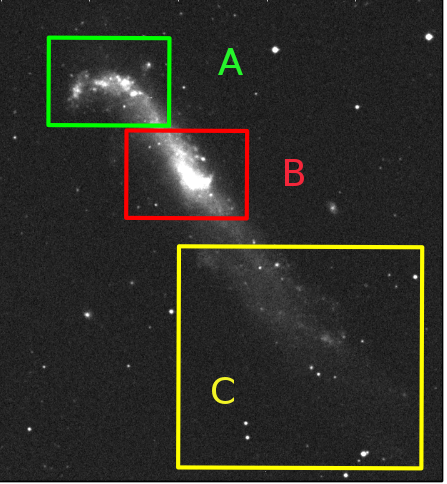} 
\caption{Digital Sky Survey R-band image for NGC 4656. Green (2.9'$\times$2.0'), red (2.9'$\times$2.0') and yellow  (5.8'$\times$5.3') boxes represent the regions that were used to compare SFRs in NGC 4656.}
%It shows the three regions in which the system has been divided (A, B and C), in order to compare the SFRs along NGC 4656. The regions belonging to the A, B and C zones are contained inside the green boxes (2.9'$\times$2'), red (2.9'$\times$2') and yellow (5.8'$\times$5.3') respectively.}
\label{3zonas_sfr}
\end{figure}

In order to determine how the SFR varies along NGC 4656, SFRs were estimated in three main sectors, namely, A, B and C, and which are shown in Fig. \ref{3zonas_sfr}. The selection of these boxes was made through visual inspection of the monochromatic map of NGC 4656 based on its morphology. We estimated the values of SFR$_{H\alpha}$=0.053$\pm$0.023 M$_{\odot}yr^{-1}$, SFR$_{H\alpha}$=0.030$\pm$0.013 M$_{\odot}yr^{-1}$ and SFR$_{H\alpha}$=0.011$\pm$0.050 M$_{\odot}yr^{-1}$ for the A, B and C regions respectively, and whose values are summarized in Table \ref{table:luminosidades}. 

Considering the large uncertainties, we note that region A displays a larger SFR than region C (despite the different covered area, see Fig. \ref{3zonas_sfr}), indicating that the north-east region of NGC 4656 is actively forming stars.
%Comparing these values derived for each field, all sectors have similar SFRs considering their uncertainties. However, considering the number of counts displayed by the monochromatic map, we found that the A region  displays the highest SFR, instead C region presents the most modest (being $\sim$2 times lower than A). 
In the case of NGC 4656 as a whole, we obtained a SFR$_{H\alpha}$=0.094$\pm$0.040 M$_{\odot}yr^{-1}$. This value of SFR$_{H\alpha}$ locates NGC 4656 in the star-forming sequence in the M$_{*}$-SFR$_{H\alpha}$ plane (considering the uncertainties), according to the relation estimated by \cite{2012zahid}\footnotemark \footnotetext{log(SFR)=0.71 log M$_{*}$-6.78 \citep{2012zahid}} (who consider a mass range of log(M$_{*}$)$\sim$8.5-10.4). For the TDG candidate, we obtained a luminosity of L$_{H\alpha}$= 4.00$\pm$2.00 $\times$10$^{38}$erg $s^{-1}$ and SFR$_{H\alpha}$=0.003$\pm$0.001 M$_{\odot}yr^{-1}$, placing this object in the star-forming sequence defined for dwarf galaxies with low-surface brightness \cite{2017mcgaugh}.

% This value of SFR$_{H\alpha}$ locates NGC 4656UV in the star-forming sequence in the plane M$_{*}$ versus SFR$_{H\alpha}$ (considering the uncertainties), according to the relation estimated by \cite{2017mcgaugh}\footnotemark \footnotetext{log(SFR)=1.04 log M$_{*}$-10.75 \citep{2017mcgaugh}} for dwarf galaxies with low surface brightness (who consider a mass range of log(M$_{*}$)$\sim$6.7-9.8). 

\begin{table}
\centering
%\resizebox{\textwidth}{!}{
\begin{threeparttable}

\caption{Luminosities and SFRs for NGC 4656 and NGC 4656UV}             % title of Table
\label{table:luminosidades}      % is used to refer this table in thSFR$_{H\alpha}$ = 0.094 M$_{\odot}yr^{-1}$e text
\centering                          % used for centering table
\begin{tabular}{lcc}        % centered columns (4 columns)
\hline\hline                 % inserts double horizontal lines
Source & L$_{H\alpha} $\tnote{\emph{ac}}  & SFR$_{H\alpha}$\tnote{\emph{bc}} \\%& L$_{H\alpha} $\tnote{\emph{ae}}  & SFR$_{H\alpha}$\tnote{\emph{be}}   & SFR$_{FUV}$\tnote{\emph{ce}} \\    % table heading 
     &$\times$10$^{40}$erg $s^{-1}$ &  M$_{\odot}yr^{-1}$ \\%&$\times$10$^{40}$erg $s^{-1}$ &  M$_{\odot}yr^{-1}$ &  M$_{\odot}yr^{-1}$ \\
\hline                        % inserts single horizontal line
   NGC 4656UV & 0.04$\pm$0.02 & 0.003$\pm$0.001 \\%& 0.08$\pm$0.02 & 0.006$\pm$0.001 & 0.027$\pm$0.001 \\      % inserting body of the table
   NGC 4656 A & 0.66$\pm$0.29 & 0.053$\pm$0.023 \\%  & 1.33$\pm$0.29 & 0.105$\pm$0.024 & - \\
   NGC 4656 B & 0.37$\pm$0.16 & 0.030$\pm$0.013 \\%& 0.74$\pm$0.16 & 0.059$\pm$0.013    & - \\
   NGC 4656 C & 0.14$\pm$0.06  & 0.011$\pm$0.005 \\%& 0.28$\pm$0.06  & 0.023$\pm$0.005   & - \\
   NGC 4656 whole & 1.18$\pm$0.51 & 0.094$\pm$0.040 \\% & 2.35$\pm$0.53 & 0.186$\pm$0.042 & 0.666$\pm$0.030 \\ 
\hline                                   %inserts single line
\end{tabular}
 \begin{tablenotes}
\footnotesize
 \item[\emph{a}]{Value estimated in this work using the integrated H$\alpha$ emission flux.}
 \item[\emph{b}]{Value estimated in this work using Kennicutt's (1998) formula}
 % \item[\emph{c}]{Valor estimado por \cite{2012schech}}
   \item[\emph{c}]{The assumed distance is 5.1 Mpc \citep{2014sorce}.}\\
 %   \item[\emph{e}]{Se ha considerado una distancia de 7.2 Mpc \citep{2005seth} a ambos sistemas.}\\
%\textbf{Nota}: Los valores presentados en esta tabla no han sido corregidos por extinción
 \end{tablenotes}
\end{threeparttable}
%}
\end{table} 
 
%NO TIENES QUE ABUSAR DEL "ON THE OTHER HAND...". ADEMAS, ESA FRASE DEJA DE TENER SENTIDO SI CONTINUAS HABLANDO DEL MISMO TEMA...

In order to compare our results with those obtained in literature, we have also estimated the luminosities and SFRs considering the distances assumed by the different authors \citep{2008pil,2010mapelli,2012schech}. The results are presented in Table \ref{table:luminosidades2}. Using a distance of 7.2$\pm$0.8 Mpc \citep{2005seth}, which was adopted by \cite{2012schech}, we obtained the value of SFR$_{H\alpha}$=0.186 M$_{\odot}yr^{-1}$ for NGC 4656 and SFR$_{H\alpha}$=0.006 M$_{\odot}yr^{-1}$ for NGC 4656UV as a whole. Based on the FUV luminosity, \cite{2012schech} found the values of SFR$_{FUV}$= 0.666 M$_{\odot}yr^{-1}$ and SFR$_{FUV}$= 0.027 M$_{\odot}yr^{-1}$ for NGC 4656 and NGC 4656UV respectively. For both cases, we found discrepancies between these last values and those calculated in this work. We can note that for NGC 4656 the SFR$_{H\alpha}$ is $\sim$3 times lower than the SFR$_{FUV}$ obtained by \cite{2012schech}, and for NGC 4656UV the SFR$_{H\alpha}$ is $\sim$4 times smaller than the SFR$_{FUV}$. Indeed, previous authors have found a systematic difference between SFRs derived from H$\alpha$ and UV emission. For example, \cite{2009lee}, found that H$\alpha$ emission tends to underestimate the total SFR with respect to FUV, in the case of low-luminosity dwarf galaxies. These authors suggest that variations in the Initial Mass Function (IMF) could explain the differences, arguing a possible deficiency of massive ionizing stars in low mass galaxies. Using a distance of 8.8 Mpc (\citealt{2008pil}), we obtained a SFR$_{H\alpha}$=0.28 M$_{\odot}yr^{-1}$ for NGC 4656. \cite{2008pil} calculated the SFR for NGC 4656 based on the L$_{FIR}$ luminosity, obtaining a value of SFR$_{FIR}$=0.23 M$_{\odot}yr^{-1}$. Comparing both SFRs we can see that both results are similar. Considering a distance of 8.7 Mpc \citep{2000mould}, value considered by \cite{2010mapelli}, we obtained the value of SFR$_{H\alpha}$=0.27 M$_{\odot}yr^{-1}$ for NGC 4656. Based also on L$_{H\alpha}$ \citep[fluxes extracted from][]{2006moustakas}, \cite{2010mapelli} estimated the value of SFR$_{H\alpha}$= 0.54 M$_{\odot}yr^{-1}$. As we can notice, \cite{2010mapelli} obtained an SFR$_{H\alpha}$$\sim$2 times higher than the one estimated in this work. A factor that can affect this discrepancy is the extinction correction, given that the attenuation by dust can cause an underestimation of the intrinsic luminosity of the system. The fluxes used in this paper were not corrected by extinction, on the other hand, the fluxes used by \cite{2010mapelli} are corrected by Galactic extinction and interstellar absorption. Another relevant factor of great influence to consider, is related to the uncertainties caused by the indirect calibration of the Fabry-Perot H$\alpha$ maps from the GMOS data (RMSE=1.32$\times$10$^{15}$ erg s$^{-1}$cm$^{-2}$, see \ref{datos_fluxcalibration}).

\begin{table*}
\centering
%\resizebox{\textwidth}{!}{
\begin{threeparttable}
\caption{SFRs for NGC 4656 and NGC 4656UV}             % title of Table
\label{table:luminosidades2}      % is used to refer this table in the text
\centering                          % used for centering table
\begin{tabular}{lcccccc}        % centered columns (4 columns)
\hline\hline                 % inserts double horizontal lines
Source   & SFR$_{H\alpha}$\tnote{\emph{ae}}   & SFR$_{FUV}$\tnote{\emph{be}} &SFR$_{H\alpha}$\tnote{\emph{af}} & SFR$_{FIR}$\tnote{\emph{cf}} &SFR$_{H\alpha}$\tnote{\emph{ag}} &SFR$_{H\alpha}$\tnote{\emph{dg}}\\    % table heading 
      &   &   &M$_{\odot}yr^{-1}$  &  & & \\%&$\times$10$^{40}$erg $s^{-1}$ &  M$_{\odot}yr^{-1}$ &  M$_{\odot}yr^{-1}$ \\
\hline                        % inserts single horizontal line
   NGC 4656UV & 0.006$\pm$0.001 & 0.027$\pm$0.001 & - & - & - & -\\      % inserting body of the table
   NGC 4656  & 0.186$\pm$0.042 & 0.666$\pm$0.030 & 0.28 & 0.23 &0.27 & 0.54 \\ 
\hline                                   %inserts single line
\end{tabular}
 \begin{tablenotes}
\footnotesize
 \item[\emph{a}]{Value estimated in this work using  the \cite{1998kenni} recipe, based on L$_{H\alpha}$}
  \item[\emph{b}]{Value estimated by \cite{2012schech}, based on L$_{FUV}$}
  \item[\emph{c}]{Value estimated by \cite{2008pil}, based on L$_{FIR}$}
  \item[\emph{d}]{Value estimated by \cite{2010mapelli}, based on L$_{H\alpha}$}
   \item[\emph{f}]{It has been considered a distance of 8.8 Mpc \citep{2008pil} to both systems.}
   \item[\emph{g}]{It has been considered a distance of 8.7 Mpc \citep{2000mould} to both systems.}\\
\textbf{Note:} Values presented on this table has not been corrected by extinction.
 \end{tablenotes}
\end{threeparttable}
%}
\end{table*}

%We derived the stellar masses M$_{*}$ for NGC 4656 and NGC 4656UV from the average mass-luminosity ratio (M/L) (whose values were 0.69 and 1.17 respectively) estimated from 9 colors based on the u', g', r, i' and z' filters, using the \cite{2003bell}  conversions. We considered the fluxes from \cite{2012schech} and the distance of 5.1 Mpc \citep{2014sorce} to both systems. For NGC 4656, 2 masses were estimated considering the luminosities in the H and K bands: $\sim$8.38$\times$10$^{8}$ and $\sim$8.76$\times$10$^{8}$M$_{\odot}$, respectively. The final stellar mass was calculated by averaging both giving a value of $\sim$8.57$\times$10$^{8}$M$_{\odot}$. For NGC 4656UV, the final stellar mass was estimated considering the luminosity in the g' band because only upper limits were estimated for the fluxes in the H and K bands. As a result, a value of $\sim$5.46$\times$10$^{7}$M$_{\odot}$ was obtained. 
%\subsection{Kinematics}

\subsection{Kinematics}
\subsubsection{Radial velocity and velocity dispersion } \label{sec:velocities}

\begin{figure}
   \centering
\includegraphics[scale=0.53]{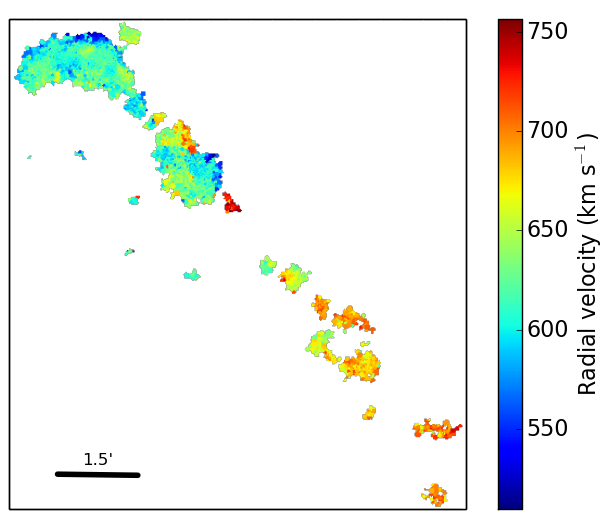} 
   \caption{Radial velocity map for NGC 4656. The color bar represents the velocity units of km s$^{-1}$}
   \label{ngc4656_vel}
    \end{figure}

Fig. \ref{ngc4656_vel} shows the velocity field along the whole NGC 4656 galaxy. On a large scale, it can be seen that the velocities from the northeast to the southwest range from blue to red respectively (from $\sim$520 to $\sim$720 km s$^{-1}$), which indicates that the global kinematic  of the galaxy is still dominated by rotation. This is nevertheless not observed when analysing the field of velocities on a smaller scale. For example, in the northeast, central and southeast areas by separate, we can see that the velocity field is very disturbed and does not present a well-behaved velocity gradient (see \S \ref{sec:curves}). 

%In Fig. \ref{ngc4656_velschech} we have superimposed on our velocity map the isovelocities contours obtained by \cite{2012schech} by observations of H{\sc i}. As expected,the radial velocity map of ionized gas is embedded within the neutral gas radial velocity map.
The velocity range found in this work for the ionized gas (from $\sim$520 to $\sim$720 km s$^{-1}$) is consistent with the velocity range for the H{\sc i} gas found by \cite{2012schech} (from $\sim$590 to $\sim$720 km s$^{-1}$). These authors found the presence of gas which is counterrotating with respect to the disk, in the northeast region of the galaxy center. In order to determine if the ionized gas presents similar characteristics, we have proceeded to analyze in detail this area considering that our data have better spatial resolution than H{\sc i} data. For this purpose have been constructed RCs, whose results are presented in \S \ref{sec:curves}.\\

%\begin{figure*}
%\centering  
 % \includegraphics[scale=0.4]{contHI_2.png} 
 %\caption{Radial velocity map for NGC 4656 on which we have superposed the HI isovelocities contours obtained by Schechtman et al. (2012). The color bar represents the velocity units of km/s} 
 %\label{ngc4656_velschech}
%\end{figure*}    

\begin{figure}
   \centering
\includegraphics[scale=0.53]{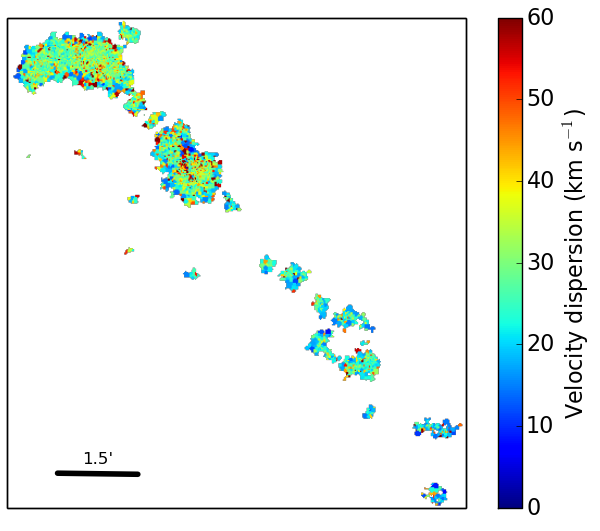} 
   \caption{Dispersion velocity map for NGC 4656. The color bar represents the velocity units of km/s}
   \label{ngc4656_disp}
    \end{figure}

%\subsubsection{Velocity dispersion maps} 
%cambiar primera linea, textua lde zasov 2016
The velocity dispersion map corrected from instrumental broadening for NGC 4656 is presented in Fig. \ref{ngc4656_disp}. It can be seen that the values of the velocities dispersion throughout the galaxy are fairly uniform ($\sigma$$\sim$24-30 km s$^{-1}$). For nearby dwarf star-forming galaxies, where the velocity dispersion reflects the inner motions of gas in H{\sc ii} regions, a velocity dispersion that does not exceed 30 km s$^{-1}$ is quite typical % I guess you should also mention Epinat 2008? (Ghasp paper) where this is done for a larger sample of galaxies.  In any case you cite moiseev 2 times, one is enough, idem for amram below... 
\citep{2015moiseev,2008epinat}. For NGC 4656, we have found a flux-weighted average velocity dispersion of $\sim$30$\pm$11 km s$^{-1}$, whose uncertainty is represented by the standard deviation. This value is within the range of typical values for dwarf star-forming galaxies. % obtained by \cite{2015moiseev}, who through Fabry-Perot data studied the kinematics of a sample of 59 nearby dwarf star-forming galaxies. 
On the other hand, for the upper part of the galaxy, the flux-weighted velocity dispersion is 29$\pm$9 km s$^{-1}$, while for the central part this value is 33$\pm$12 km s$^{-1}$  and for the tail 26$\pm$12 km s$^{-1}$.\\%mean gas velocity dispersion of 29.17 km/s, flux wighted velocity 30.13 . 

Non-virial gravitational motions such as ISM turbulence produced by tidal interactions, galaxy mergers and gas accretion events, can influence in the broadening of emission lines that are observed in an integrated way in galaxies \citep{2011bournaud,2014arribas}. For this reason, we also analyze the velocity dispersion in the area that connects the two possible galaxies (NGC 4657 and NGC 4656), which was previously presented in Fig. \ref{perfiles_horizontal}.
%Fig. \ref{inter_disp} shows the velocity dispersion map corresponding to this specific region, with the H$\alpha$ profiles overlapped. 
 This region is not dominated by wide profiles ($\sigma$ greater than 30 km s$^{-1}$), being the flux-weighted velocity dispersion of $\sim$27 km s $^{-1}$. The presence of wide profiles would indicate the presence of multiple components due to a merger process between two galaxies (NGC 4657 and NGC 4656), which would mean that the two systems still have their own kinematic and that they are not totally relaxed. An example is the work done by \cite{2007amram} for the case of HCG 31, where they find the presence of wide profiles greater than 30 km s$^{-1}$ in the region that connects galaxies A and C, evidence of an ongoing merging. \\

%\begin{figure}
%\centering
%\includegraphics[scale=0.3]{inter_disp+prof.png} 
%\caption{Velocity dispersion map for the region shown in \ref{perfiles_horizontal} (blue square, top panel). The H$\alpha$ emission profiles normalized to the brightest profile have been superimposed. OLD}
%\caption{Velocity dispersion map for the region shown in \ref{perfiles_horizontal}. The H$\alpha$ emission profiles normalized to the brightest profile have been superimposed.}
%\label{inter_disp}
%\end{figure}

%\subsubsection{RCs: Classical method} \label{sec:curves}       
\subsubsection{Rotation Curves} \label{sec:curves}   
%Classical method:\\

\begin{figure}
 \hspace{-0.5cm}
 \includegraphics[scale=0.53]{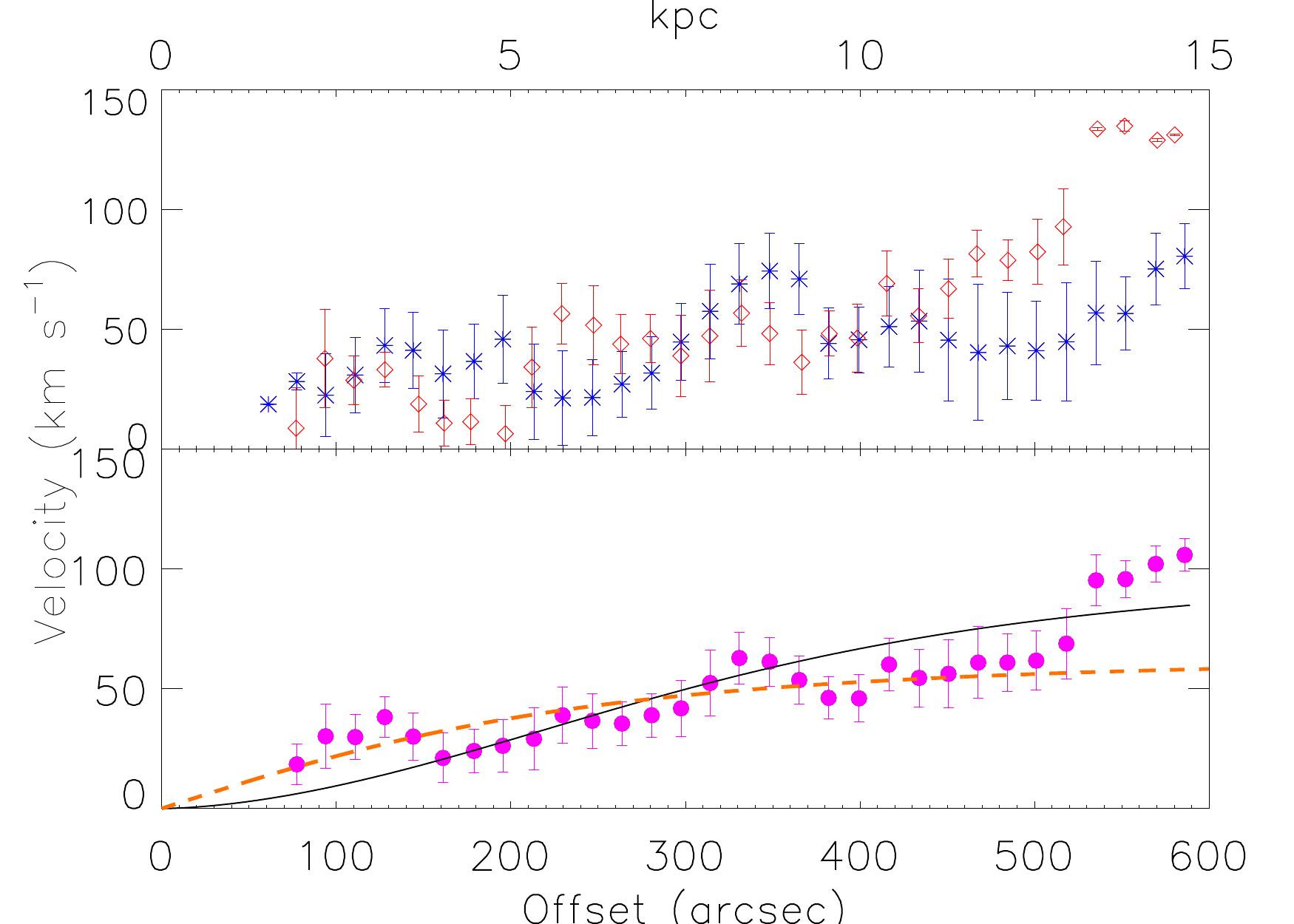}
 \caption{Rotation curves for NGC 4656. Top panel: Curve parameters are PA=40$^{\circ}$, i= 82$^{\circ}$ and V$_{sys}$= 652 km s$^{-1}$. Asterisks and diamonds represents the average value of velocities contained in crowns of 25 pixels. Red symbols represent the receding side and blue symbols represent the approaching side of the curve. The associated error bars correspond to the mean deviation. Bottom panel: curve obtained by the average between the approaching and the receding side from the curve presented in the top panel. The continuous black line represent the Zhao function \citep{1998krav} fitted to the points. The orange dashed line represents the RC from the best fitting providing a logarithm-like potential (see \S \ref{results_model} for more details).} % I SUGGEST TO WRITE INSTEAD: The orange dashed line represents the RC from the best fitting providing a logarithm-like potential (see \S \ref{results_model} for more details).\color{black}} (instead of the \citet{1991bertola}'s model)
\label{curvas_whole}   
\end{figure}

We have derived RCs for the whole galaxy NGC 4656, with different parameters. In both panels of Fig. \ref{curvas_whole}, we present the best RC obtained, whose parameters are listed  in Table \ref{table:parametros_curva}. The inclination and position angle of the major axis were fixed along the radius, given that the line map and velocity field do not indicate the presence of warps or flares, except at the north-east region of this system (where ``NGC 4657'' is located).  The top panel presents the approaching (blue symbols) and receding (red symbols) sides of the curve. Asterisks and diamonds represent the average value of velocities contained in crowns of 25 pixels. The bottom panel presents the average curve obtained from the approaching and receding sides shown in the top panel. A fit represented by the continuous black line is showed, which corresponds to the Zhao function \citep{1998krav}. From this model, we obtained that the maximum rotational velocity of the galaxy is 83 km s$^{-1}$, at a radial distance of 600 arcsec of the kinematic center.

%In order to estimate the maximum rotational velocity of NGC 4656, from the curve shown in the lower panel of Fig \ref{curvas_whole} we have constructed a new curve in which we have averaged the velocities of the approaching an receding side, and then we made a fitting using the function described by  Kravtsov et al. (1998). The result is shown in Fig \ref{curva_promedio}. From this model, we obtained that the maximum rotational velocity of the galaxy is 83 km s$^{-1}$ (this value and other parameters of the curve, are listed in Table \ref{table:parametros_curva}).\\

%\begin{figure}
%\centering
%\includegraphics[scale=0.53]{curvarot_promedio_ajuste.pdf}
%\caption{Rotation curve for NGC 4656 obtained by the average between the approaching and the receding side from the curve presented in the lower panel of Fig \ref{curvas_whole}. The continuous black line represent the fit.}
%\label{curva_promedio}
%\end{figure}

\begin{table} \caption{Parameters for the rotation curve of NGC 4656}\label{table:parametros_curva}
\centering
%\begin{center}
%\resizebox{0.5\textwidth}{!}{
\begin{threeparttable}
\begin{tabular}{ll}
%\caption[]{Propiedades principales de la muestra}
\hline\hline 
Parameter &Value\\ \hline \\
Rotation center\tnote{\emph{a}} &  R.A. 12$^{h}$43$^{m}$52.893$^{s}$\\
 & Dec.
  32$^{d}$08$^{m}$54.56$^{s}$\\ 
  Inclination\tnote{\emph{b}} & 82$^{\circ}$ \\
  PA\tnote{\emph{c}} & 40$^{\circ}$ \\
 Systemic Velocity \color{black}\tnote{\emph{a}} &  652 km s$^{-1}$\\
 Maximum rotational velocity\tnote{\emph{a}} & 83 km s$^{-1}$  \\
 \hline
\end{tabular}
\begin{tablenotes}
\footnotesize
\item[\emph{a}]{Values estimated in this work.}
\item[\emph{b}]{Value extracted from \cite{1983stayton}}
\item[\emph{c}]{Value extracted from \cite{2012schech}}
\end{tablenotes} 
%\end{threeparttable}
%\end{center}
\end{threeparttable}
%}
\end{table}

%%%%%%%%%%cambiar de lugar
The nature of the gravitational support of a system in equilibrium is given by the ratio of the maximum circular rotation velocity V$_{max}$ and the local velocity dispersion $\sigma$.  Due to gravitational perturbation due to interaction with the companion galaxies, NGC 4656 is maybe not a system at equilibrium, so only as a reference we have calculated that ratio.
Considering the previously obtained values V$_{max}$$\sim$83 km s$^{-1}$ and $\sigma$$\sim$30 km s$^{-1}$, the ratio between both values is V$_{max}$/$\sigma$$\gtrsim$2.8. In this case, we have a high circular velocity compared to velocity dispersion (V$_{max}$/$\sigma$$>$1) which is the signature of a rotation-dominated gravitational support.\\ %Vmax is due to local motion, so it could not been used to estimate the gravitational support. I propose that you use instead a model to determine Vmax. You can use for instance a Zhao function of any function to estimate Vmax and we did in Epinat et al.’s paper. The advantage of the Zhao function is its high number of freee parameters. Vt, Rt, g, a and b are free parameters, so with this 5-parameters curve you can fit what ever you want ! Please force the fit to pass by (0.0). But you can use a more simple fit if you prefert, arctan, and so on P.AMRAM

\begin{figure*}
   \centering 
  \includegraphics[scale=0.33]{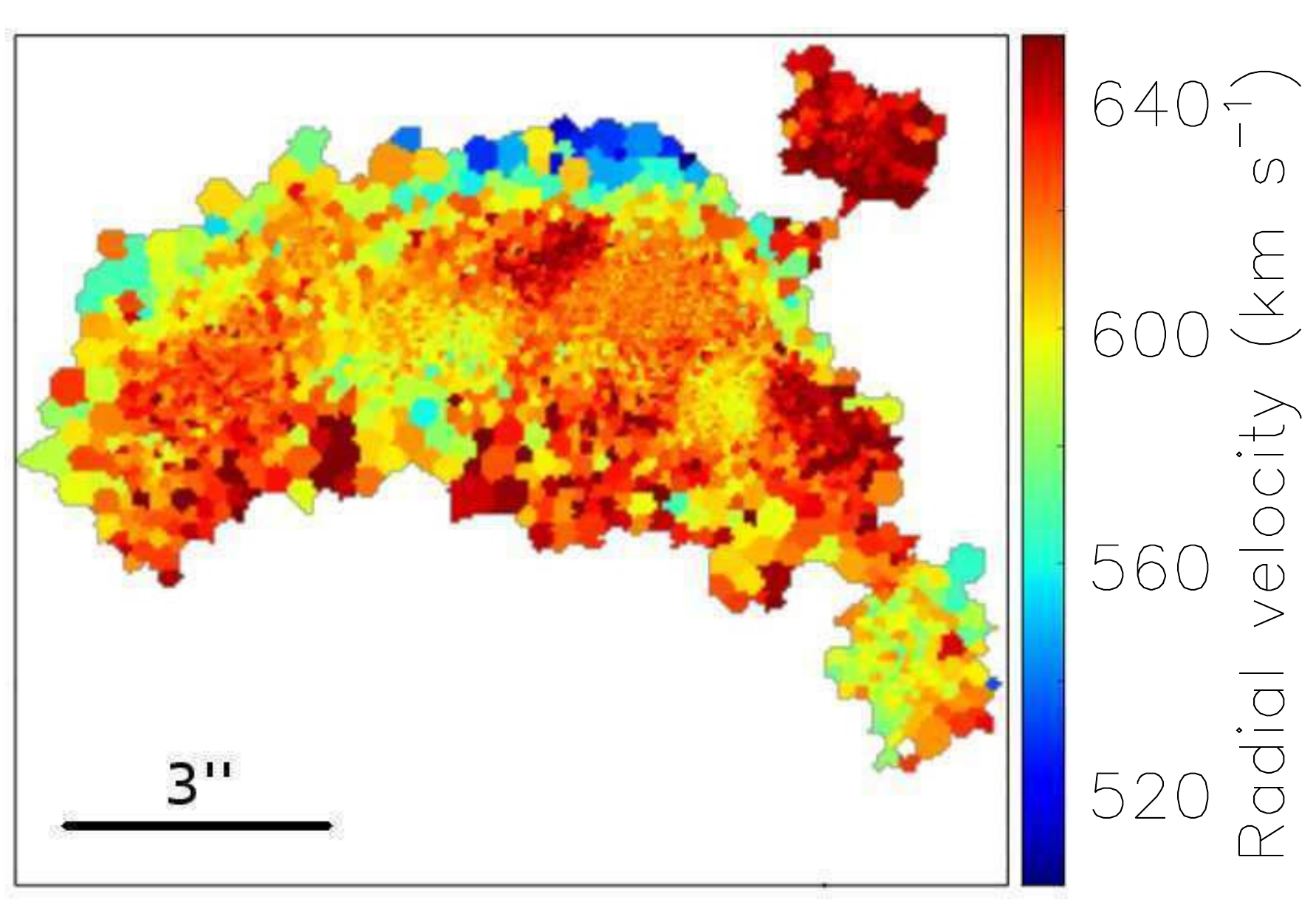}
  \includegraphics[scale=0.6]{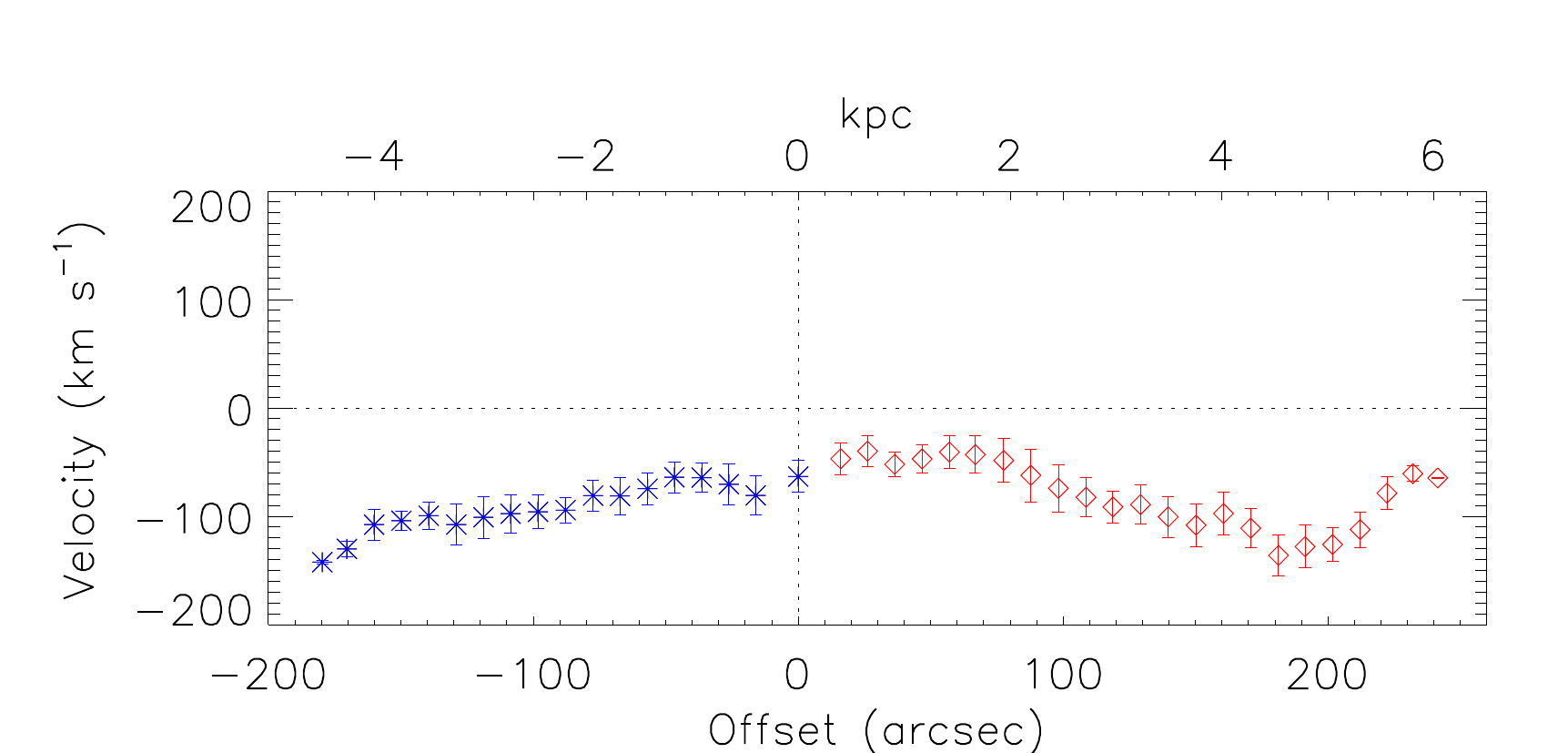}
  \includegraphics[scale=0.33]{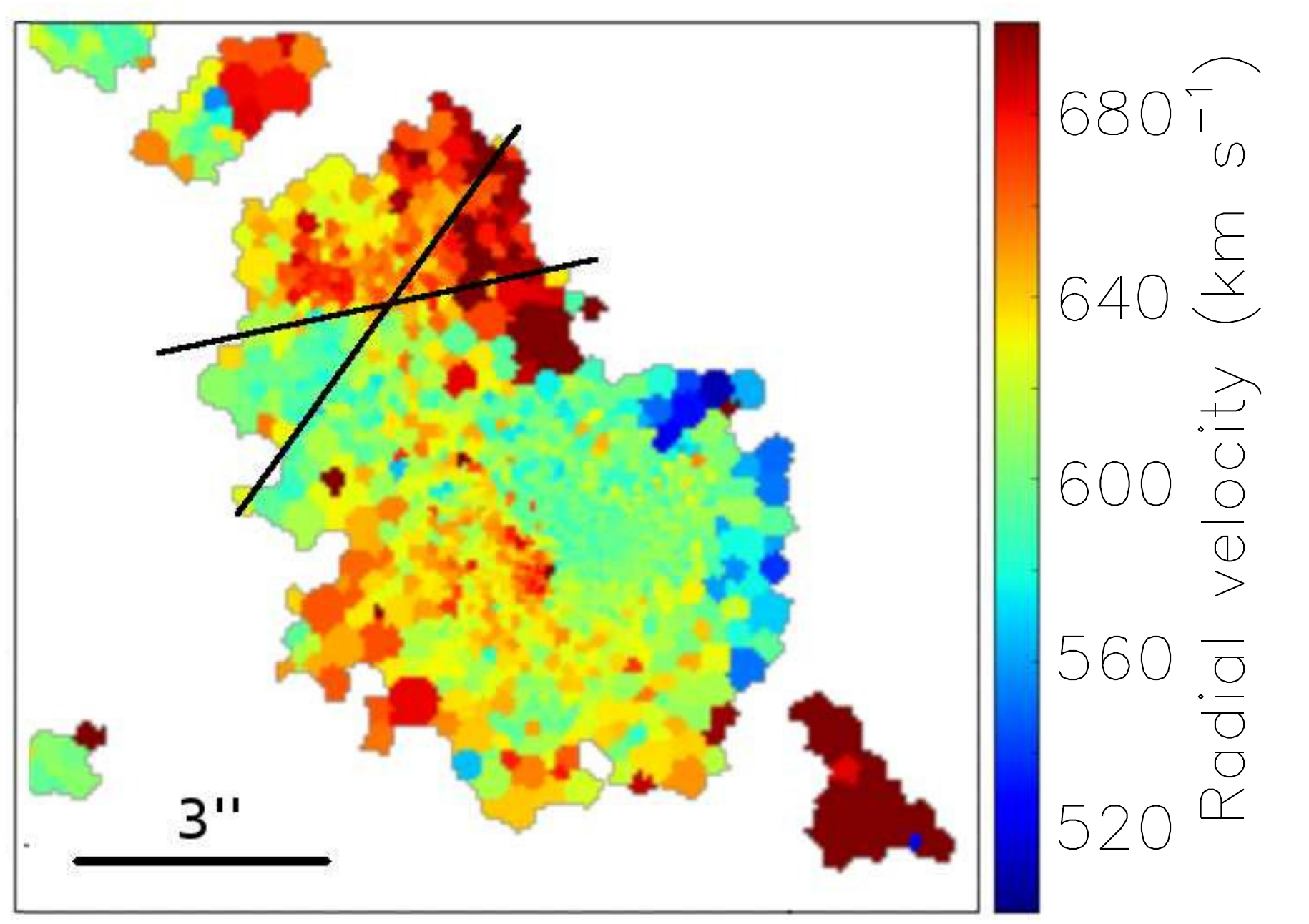}
 \includegraphics[scale=0.6]{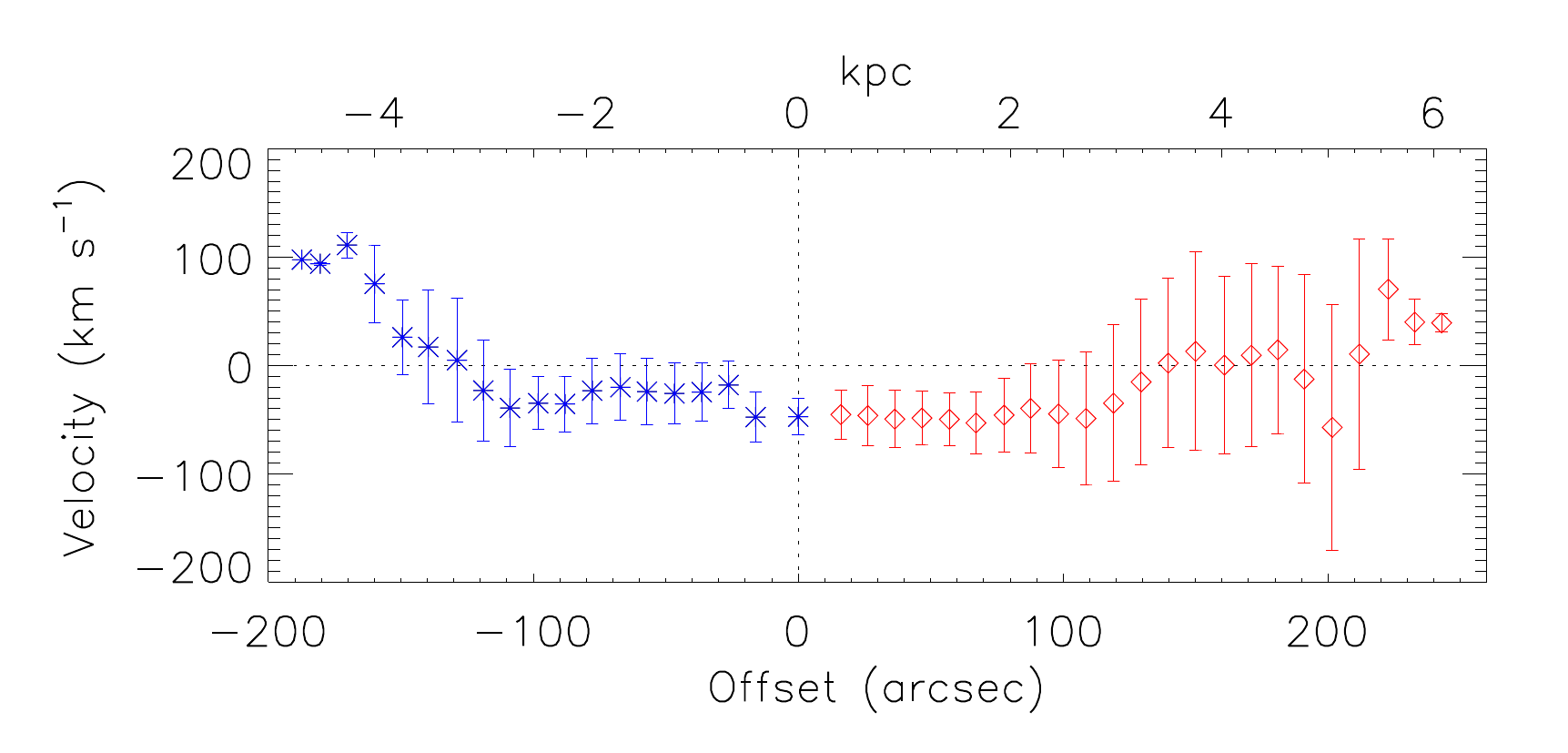}

 \caption{Top left panel:  Velocity map for NGC 4657. Top right panel: Position-velocity plot for NGC 4657, constructed from the map shown in the top left panel. The parameters used are PA = 70$^{\circ}$, i = 82$^{\circ}$ and V$syst$ = 652 km s$^{-1}$. An angular sector of 20$^{\circ}$ has been chosen around the major PA. Bottom left panel: Velocity map for the center of NGC 4656.  Two crossed black lines have been superimposed over an area that has a velocity gradient, which represent two cones that contain the points used to make the rotation curves shown in Fig. \ref{curvas_counter}. The intersection between both curves represents the center used for the curves. Bottom right panel: Position-velocity plot for the center of NGC 4656, constructed from the map shown in the bottom left panel. The parameters used are PA = 35$^{\circ}$, i = 82$^{\circ}$ and V$syst$ = 652 km s$^{-1}$.  An angular sector of 30$^{\circ}$ has been chosen around the major PA.}
  \label{vel_56y57}
\end{figure*}

%Another way to look for evidence of an interaction between NGC 4656 and NGC 4657, is to study the continuity in the kinematic between both systems. 
In order to investigate if NGC 4657 could be self-gravitating we zoom its velocity field in the top left panel of Fig. \ref{vel_56y57}. We performed the same analysis on the center of NGC 4656 (Fig. \ref{vel_56y57} bottom left panel) that also looks disconnected from the rest of the system. The left panels of Fig. \ref{vel_56y57} shows that both systems display no clear velocity gradients (a sort of velocity gradients can be seen at very small physical scales), which is insufficient to indicate that NGC 4657 is dominated by circular motion, except of course if the system is almost face-on, which does not seem to be the case.
%therefore are not dominated by circular motions. %On the other hand, the lowest velocities in both systems (blue bins) are located on the edges of the velocity fields, they are large (meaning that their SNR is rather low) and are few (with respect to the others bins ranging from light green to dark red). 
In addition to the low amplitude velocity gradient, the perturbed velocity fields do not allow to compute any reliable RC, considering that in addition the inclination for both system are impossible to evaluate due to their peculiar morphology. As example, the right panel of Fig. \ref{vel_56y57} shows two construction attempts of the RC for NGC 4657 and the center of NGC 4656. %Considering that it is not possible to estimate a reliably rotation curve, we plot position velocity-plot along both systems as attempted in the left panel of Fig. \ref{vel_56y57}. We make those plots using position angles at xxx and xxx in order to %(I do not understand why you chose those angles by the way). 
Those velocity plots show an almost constant radial velocity along the cut and confirm that none of these entities are self-gravitating.
%The blue pixels observed in Fig. 16 is clearly seen in the blue tail of NGC 4657. %Please replot the graph and then we will do the additional comments on these plots. 

%\begin{figure*}
  % \centering
  %  \includegraphics[scale=0.3]{NGC4657_vel.pdf}\\ 
   % \includegraphics[scale=0.3]{NGC4656centro_vel.pdf} 
  % \includegraphics[scale=0.3]{fig2.png}
   % \caption{Top panel: Velocity map for NGC 4657. Bottom panel: Velocity map for the center of NGC 4656.}
  %  \label{vel_56y57}
%\end{figure*}   

%\begin{figure}
 %  \centering
%\includegraphics[scale=0.53]{curvasfallidas_paper.pdf-converted-to.pdf}
%\caption{Top panel: Rotation curve for NGC 4657, constructed from the map shown in the top panel of Fig. \ref{vel_56y57}. The parameters used are PA = 70$^{\circ}$, i = 82$^{\circ}$ and V$syst$ = 652 km s$^{-1}$. Bottom panel: Rotation curve for the center of NGC 4656, constructed from the map shown in the bottom panel of Fig. \ref{vel_56y57}. The parameters used are PA = 35$^{\circ}$, i = 82$^{\circ}$ and V$syst$ = 652 km s$^{-1}$.}
%\label{curvas_fallidas}
%\end{figure}   

%

\begin{figure}
%\centering
%\includegraphics[scale=0.46]{curvascountzone_paper2.pdf-converted-to.pdf}
\hspace{-0.5cm}
\includegraphics[scale=0.52]{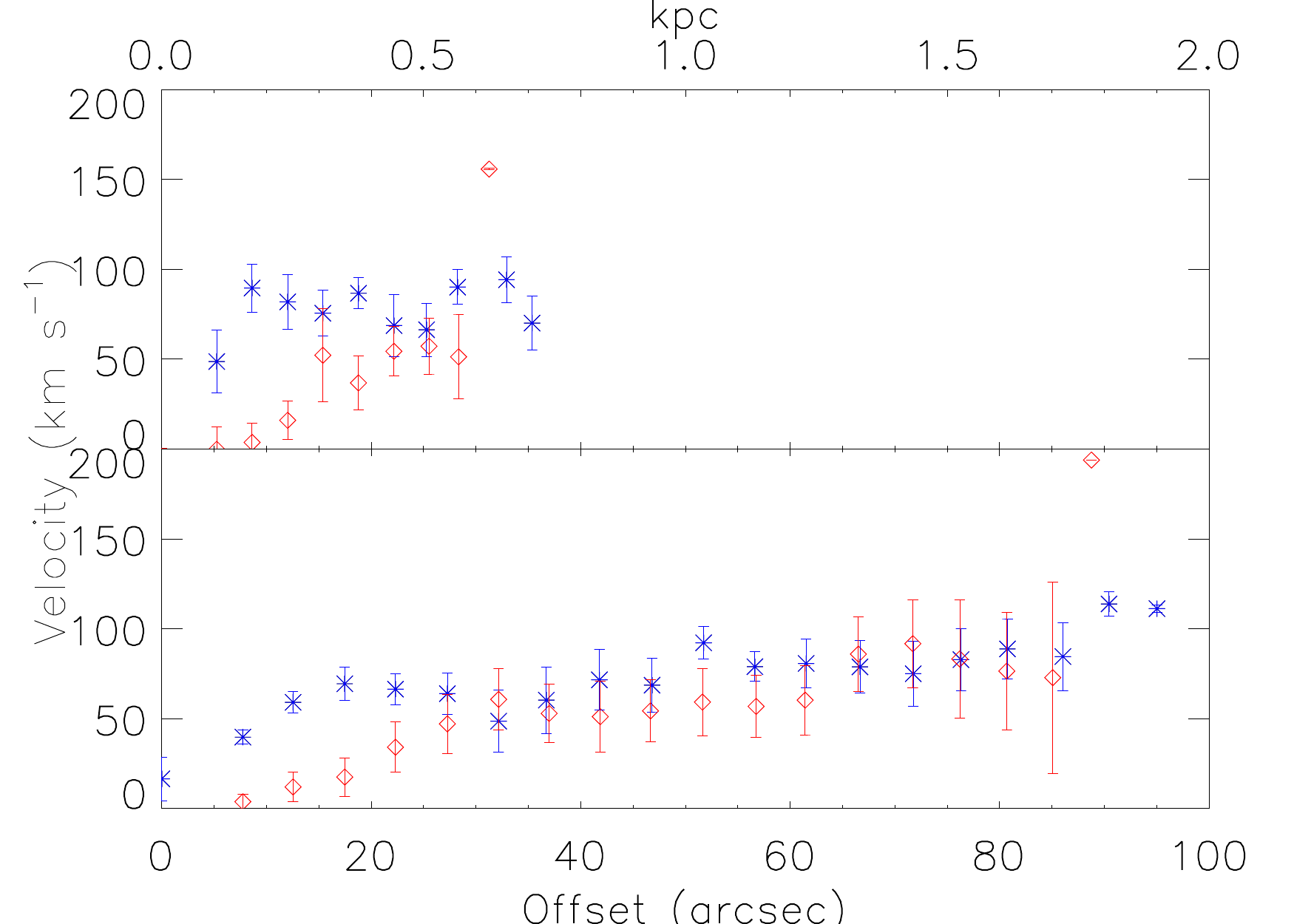} 
\caption{Rotation curves for the counter-rotating zone indicated by the black cones in Fig. \ref{vel_56y57} bottom left panel. Top panel: The parameters of the curve are PA = 130$^{\circ}$, i = 40$^{\circ}$ and V$_{syst}$ = 645 km s$^{-1}$. The left panel shows the curve along the entire extension of the region, considering as center equal to 0 arcsec, while the right panel shows the superposition of both sides of the curve in a single quadrant. Bottom panel: The parameters of the curve are PA = 130$^{\circ}$, i = 80$^{\circ}$ and V$_{syst}$ = 645 km s$^{-1}$. The left and right panel represent the same as in the top Panel.}
 \label{curvas_counter}
\end{figure}

On the other hand, analyzing the velocity maps in detail we can notice that there is a small area in the northeast sector of the center that has a rather large velocity gradient, which could be related a local rotation of gas which can be associated with an independent kinematic underlying a self gravitating structure. To study this possibility, rotation curves were produced in this area, considering the regions covered by the cone delimited by the two crossed lines in bottom left panel of Fig. \ref{vel_56y57}. %(I do not think by the way that it desserve a new plot ! you can put the sector drawn here in figure 16). The main problem of this approach is that you do not know the inclination of this structure. P.Amram
As an example, two rotation curves obtained considering two different inclination are presented in Fig \ref{curvas_counter}. %80 degrees probably prodived an upper limit for the inclination and a low rotational amplitude (Vmax ), 40 deg rees provide an higher  estimation of Vmax. See note in pdf.  P.Amram
In spite of the different inclination values used, the RC is relevant since a reasonable agreement is obtained between the curves of the approaching and receding sides. We can furthermore consider that this region is rotating with a position angle almost perpendicular to the galaxy as a whole. This is according to the findings of \cite{2012schech}, who see gas north-east of the center of the galaxy which is counterrotating with respect to the disk.

%\begin{figure}
%\centering
%\includegraphics[scale=0.22]{NGC4656centro_veledit.pdf-converted-to.pdf}
%\caption{Velocity map presented in Fig. \ref{vel_56y57} (bottom panel), with the difference that in this figure two crossed black lines have been superimposed over an area that has a velocity gradient. These lines represent two cones that contain the points used to make the rotation curves shown in Fig. \ref{curvas_counter}. The intersection between both curves represents the center used for the curves.}
%\label{mapasvel_56cono}
%\end{figure}

%\subsubsection{Gravitational support}
%\subsubsection{RCs: PV diagrams}\label{sec:resul_pv}
\subsubsection{Position-Velocity Diagrams}\label{sec:resul_pv}
\begin{figure}
\hspace{-0.5cm}
\includegraphics[scale=0.52]{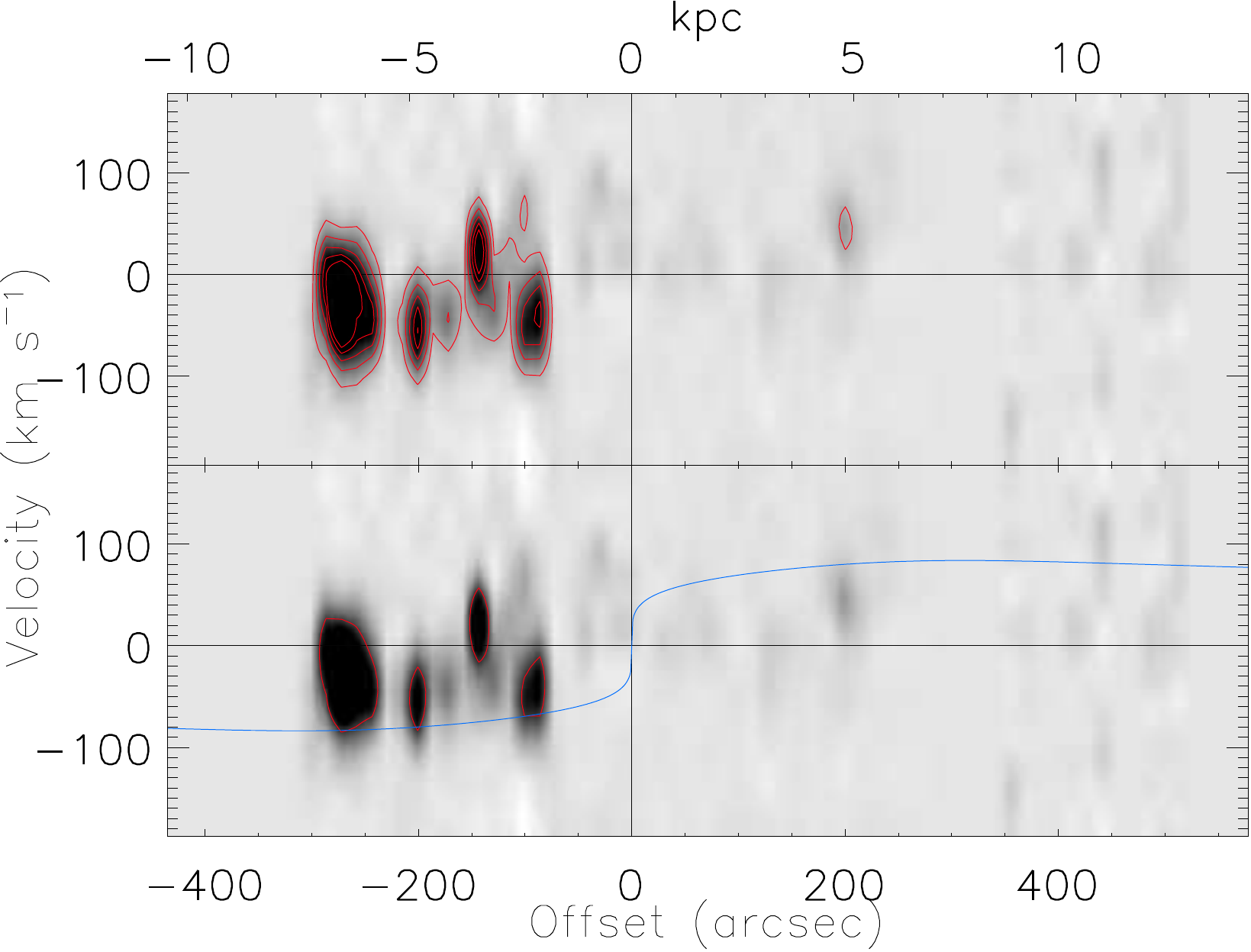}
\caption{PV diagram for NGC 4656, aligned with the major axis of the system. Top panel: the blue isocontours represent the H$\alpha$ flux at 1, 3, 5, 7 and 9$\sigma$ over the RMS noise. Bottom panel: The red isocontour represents the envelope defined at 5$\sigma$ above the RMS noise. The continuous blue line represents the Zhao fitting made on the envelope.}
\label{pvs}
\end{figure}

The Position-Velocity (PV) diagram for NGC 4656 along the major axis of the galaxy is shown in the upper panel of Fig. \ref{pvs}, where the isocontours represent the intensities at 1, 3, 5, 7 and 9$\sigma$ above the RMS noise of the PV diagram.%, where $\sigma$= .

Taking into account the broadening of the H$\alpha$ profiles due to the integration of these along the line of sight, %(spectra belonging to different regions in the plane of the disk) \
that tends to provide a rotation curve with a solid body shape instead of a more flatten one, at each radius the line-of-sight velocity is provided by the one with the largest deviation from the systemic velocity, after correction for instrumental broadening and random motions.   

We estimated  an external limit for the PV diagram from the outer envelope of the PV diagram. This envelope is defined as the isocontour corresponding to 5$\sigma$ over the RMS noise, after making a correction for the average velocity dispersion of the system ($\sim$30 km s$^{-1}$, see \ref{sec:velocities}), in order to avoid influence of local motions. Similar to the procedure described by \cite{2005garrido}, we fitted a function of Zhao \citep{1998krav} which is shown in the lower panel of Fig. \ref{pvs} (continuous blue line). From this fitting,  we compute a function that is flatter and that represents the actual LoS velocity of this almost edge-on galaxy; in addition we derived the maximum rotation velocity as V$_{max}$=V$_{Rmax}$/$sin$($i$), where  V$_{Rmax}$  corresponds to the maximum radial velocity of the envelope, and $i$ to the inclination  \citep[82$^{\circ}$,][]{1983stayton}.
Finally, we obtained the value of  V$_{max}$=85 km s$^{-1}$, which is similar to the one obtained with the rotation curve constructed through the classical method  (83 km s$^{-1}$). Considering this new value, we obtain the same previous ratio of V$_{max}$/$\sigma$$\gtrsim$2.8, which confirms the gravitational support of NGC 4656 is dominated by rotation. It is worth to note that the expected maximum rotational velocity for NGC 4656 from the near infrared Tully-Fisher relation \citep{2011torresflores} corresponds to 92$\pm$18 km s$^{-1}$ (considering an apparent magnitude of 9.1 and a distance modulus of 28.5). This value is fully consistent with our estimates of V$_{max}$ for this system.

\subsubsection{Velocity field model} \label{results_model}

\begin{figure*}
\centering \includegraphics[width=\textwidth]{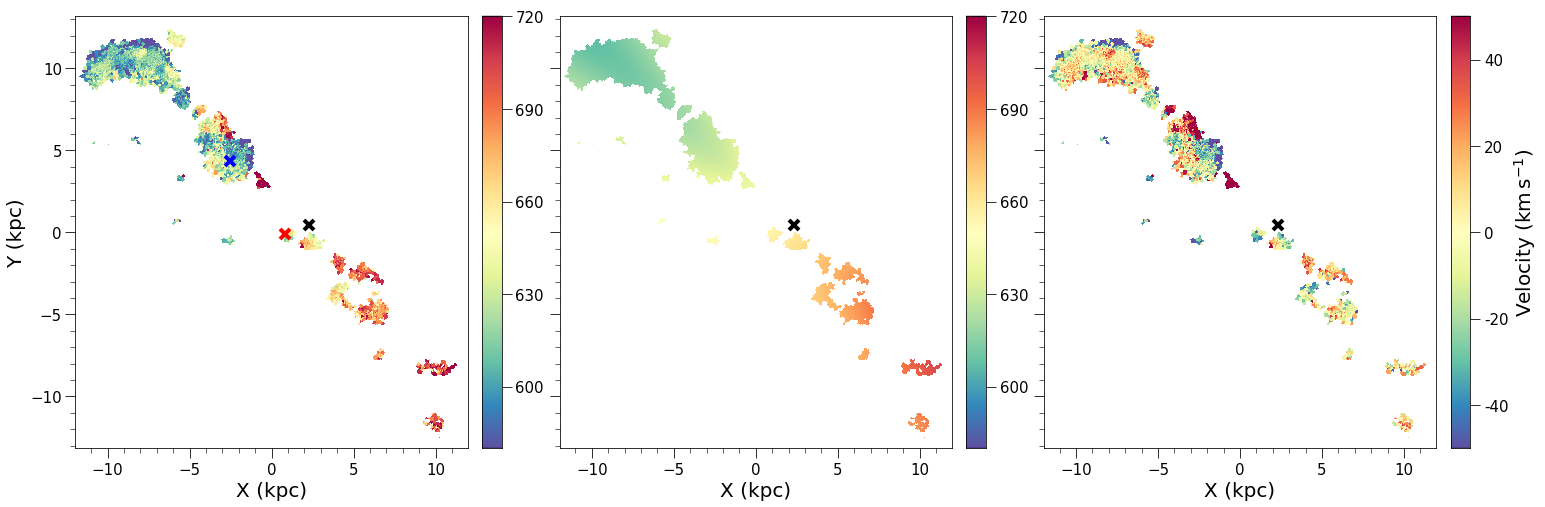}
\caption{Kinematic model of NGC4656. Left: observed velocity field in H$\alpha$ of NGC4656. Center: the best  model fitted. Right: the residual map of velocities. The photometric center of the galaxy is marked with the blue ``x'', The kinematic centers in H$\alpha$ and
HI \citep[obtained from ][]{2012schech} are marked with the black and red ``x'' respectively.}
\label{model} 
\end{figure*}

The Fig. \ref{model} shows the observed velocity field (left panel), the fitted kinematic model (center panel), and the residual map (right panel).  Following \cite{1973warner} or \cite{1978vanderkruit}, we do not observe any structure in the residual velocity field that could be the consequence of an incorrect determination of the projection parameters (inclination, centre, position angle, systemic velocity). Streaming motion that may occur in such an interacting system are not observed in those maps.  Instead, residual velocities are mainly linked to expanding bubble due to star formation events. The best fitted parameters and their respective uncertainties are given in Table ~\ref{table_model}, as well as the reduce $\chi^2$ of the fitting. We can see from the residual map that for bulk of points the fit was rather good. In fact, nearly 70\% of the profiles have a rms of 13.7\,km\,s$^{-1}$, which is quite consistent with uncertainties. Most profiles with high residuals are localized in two regions at the  west edge-border of the central part of NGC\,4656 (compare the zoom of the velocity field at left-bottom panel of Fig. \ref{vel_56y57} and the residual image of Fig. \ref{model}), even the north one is part of  counter-rotating zone identified  in \S \ref{sec:curves}. The kinematic center given by the fitting is in a good agreement, within the spatial resolution, with that found by \citet{2012schech} for H{\sc i} data (red ``x'' at the left panel of Fig. \ref{model}).
In bottom panel of Fig.~\ref{curvas_whole}, we plot the RC of
the best-fitting model (orange dashed line) overlaid on average curve between approaching and residing sides. It can be seen that there is 
a good match between the RC and the observations.
The RC reaches its peak around $12.1\pm1.3$\,kpc, becoming flattened from this point. 
This form of the RC is typical of logarithm-like potentials ($p=1\pm0.1$), as  it is expected for a disk galaxy as NGC\,4656. We can compute a dynamical mass
for NGC\,4656, at the radius (R=12.1\,kpc) of the RC peak, $M(R)=\alpha \frac{RV^{2}(R)}{G}$ where $\alpha$ accounts for the flatness of the disk.  By assuming a flat disk model, i.e. $\alpha = 0.6$ \citep{1973nordsieck}; the estimated value for dynamical mass at this radius is $6.8^{1.8}_{-0.6}\times10^{9}$\,M$_{\sun}$. The rotational velocities being quite lower than $\sim$40-60 km s$^{-1}$ up to a large radius of $\sim$ 12 kpc, it is likely that, in addition to rotation, part of the gravitational support is provided by velocity dispersion. Taking into account that $\alpha$ grows up to 1 when the infinitely flat disk component thickenses up to a spherical structure, this mass is a lower limit.   

%projected velocity profile
%(with a PA=40$^{\circ}$ and the same kinematic center given
%in Table~\ref{table:parametros_curva})

\begin{table} \caption{The fitted parameters of the kinematic model.}
\label{table_model}
\centering
\begin{threeparttable}
\begin{tabular}{ll}
\hline\hline 
Parameter             & Values \\ \hline
$V_0$    & $64\pm41$\,km s$^{-1}$ \\  
$c_{0}$    & $12.1\pm1.3$\,kpc  \\  %(275$\pm$ 30 arcsec)
p                     & $1.0\pm0.1$   \\ 
v$_{s}$  & $658\pm1$\,km s$^{-1}$ \\ 
$\psi_{0}$            & $47.5\degr\pm0.4\degr$ \\ 
$R_{ra}$      & 12:43:48.823\\ 
$R_{dec}$       & +32:08:32.88 \\ \
$\chi_{red}^{2}$       & 11.9\\ 
$M_{dyn}(R=12.1\,kpc)$ & $6.8^{1.8}_{-0.6}\times10^{9}$\,M$_{\sun}$\\ 
\hline
\end{tabular}

\end{threeparttable}
\end{table}

%%%%%%%%%%%%%%%%%%%%%%%discusion%%%%%%%%%%%%%%%%%%%%%%%%%%%%%%%%%

\section{Discussion} \label{sec:discussion}

\subsection{NGC 4656: one or two entities?}

According to \cite{2011maka}, NGC 4656 is immersed in a group of galaxies that has 28 members in a harmonic radius of 243 kpc, being NGC 4631 the brightest galaxy. In this enviroment, it is common for interactions to occur between their members \citep{1994bothun}.

Some authors considered that NGC 4656 is actually composed by two galaxies: NGC 4657, which would correspond to the northeast extreme of the system, and NGC 4656 that would correspond to the southwest sector, including the optical center \citep{1964vaucouleurs, 1973nilson}. For example, according to \cite{1964vaucouleurs}, NGC 4656 and NGC 4657 would be in the process of merger, analogous to The Antennae, and on the other hand, \cite{1973nilson} propose that the continuity between both galaxies would be only a projection effect, with NGC 4657 superimposed on NGC 4656.

In this paper, we performed a detailed analysis devoted to determine the nature of the NGC 4656/57 system.%, that is, if it corresponds to one or two galaxies.
Through the construction of rotation curves by the classic method and PV diagrams, we find that NGC 4656 is dominated by rotation  (V$_{max}$/$\sigma$$\gtrsim$2.8) on a large scale. The rotation remains predominant but disturbed at smaller scales. Furthermore, as can be seen in the Fig. \ref{model} the whole observed velocity field is consistent with the projected phenomenological model that assume only rotation motions in a plane \citep{1991bertola}. On the other hand, no independent rotations are observed for the northeast and southwest regions of the system, which if observed, could confirm the existence of two individual entities (as suggested by previous authors). %mean the presence of two dwarf galaxies in fusion, with different rotational motions (considering that both galaxies are sustained by rotation).
We also analyzed the H$\alpha$ profiles in the area that connects the two possible galaxies, since the presence of double emission profiles could be associated with star forming processes associated with different galaxies. We found the presence of some asymmetric and double peaked profiles, however these have a low emission and there is no progressive overlap of the profiles in the whole region, as found by \cite{2007amram} for the central region of the merging system HCG 31. Therefore, in the case of NGC 4646, these profiles would not indicate a merger between two galaxies, concluding that NGC 4656 is a single perturbed galaxy. 

%\textbf{Me encanto este parrafo Nahir.}

\subsection{Oxygen abundance gradient}

The distribution of metals in the disks of spiral galaxies has been studied by several authors \citep{1994zari,1998vanzee,2010moustakas,2014sanchezb}, who have found that most of spiral galaxies show clear chemical gradients, where the central regions are most metal rich than the outskirts. On the other hand, numerous observational studies have found that interacting galaxies have flatter gradients and lower nuclear metallicities than isolated galaxies \citep{2010rupke_b,2010kewley, 2012bresolin, 2012rich, 2014rosa, 2014torres, 2015olave}. Numerical simulations indicate that this drop in nuclear abundances is produced by gas inflows and gas redistributions along the galactic discs, which also produces a general flattening in the metal distribution  \citep {2010rupke_a, 2010montuori,2012torrey}.

For the case of irregular dwarf galaxies, numerous observational studies show that, in general, these galaxies have a low or no oxygen abundance gradient \citep{1999hunter,2006vanzee,2009croxall}. However, \cite{2015pilyuginb} found that irregular galaxies can also exhibit pronounced oxygen abundance gradients, there being a strong correlation between this radial gradient and the surface brightness profile, in this way galaxies with a flat brightness profile would show a gradient of soft or zero oxygen abundance, and vice versa. These authors indicate that irregular galaxies with a pronounced brightness profile would not be under the influence of a strong radial mixture of gases, on the other hand irregular dwarf galaxies with a flat brightness profile would be affected by this mixture of gases (for example, through radial gas flows), resulting in flatter abundance gradients compared to those galaxies with pronounced brightness profiles.

%due to the fact that the gas components in the galactic discs mix and redistribute, causing the gas from the outskirts (of lower metallicity) to move towards the interiors (greater metallicity), which at the same time produces flatter abundance gradients \citep {2010rupke_a, 2010montuori,2012torrey}. 

%In general, the non-radial forces exerted by one of the galaxies in the disc of its companion cause that part of the gas loses angular momentum and flows towards the center where it can generate a burst of star formation \citep{2008dimatteo}, and another part of the gas gains angular momentum and leaves the disk in long tidal tails. \cite{2012torrey} suggest that nearby galactic passages lead to low-metallicity gas flows that are quickly followed by star formation, producing in this way a small enhancement in the chemical enrichment of the medium. However, it is unlikely that solely the star formation triggered in the interactions can produce enough metals to enrich the medium to the values currently observed along the entire discs of interacting galaxies \citep{2012bresolin}. 

Towards the northeast region of the optical center of NGC 4656, %which corresponds to the brightest sector of this galaxy, 
the oxygen abundance gradient displays a slope of $\beta$=-0.027$\pm$0.029 dex kpc $^{-1}$, which suggests that it is almost flat, considering the uncertainties. In the case of the southwest region of NGC 4656, which is shaped like an extended tail, it was not possible to measure the chemical abundances with the N2 calibrator of \cite{2013marino}, since the H{\sc ii} regions have an N2 index outside the valid range for this calibrator (-1.6$<$N2$<$-0.2), which means that its oxygen abundances have a value less than 12+log(O/H)$\sim$8.0 (see Fig. \ref{abundancias}, top panel). Therefore, the distribution of metals across NGC 4656 is not symmetric with respect to its center. This kind of chemical azimuthal variations are not common, and just a few galaxies have shown features associated with these azimuthal variations \citep[e.g.,][]{2017ho,2018ho}. In addition, these chemical disturbances have not been widely studied for interacting galaxies. In this context, the rotation motions play a fundamental role in gas mixing across the galaxy disks \citep{1995roy}. For these reasons, we have estimated the time that NGC 4656 would take to complete an entire rotation. Assuming a rotational velocity of 85 km s$^{-1}$, we estimate a value of $\sim$800 Myrs. This time is much longer than the time scale associated with the interaction event that take place in NGC 4656 ($\sim$200-300 Myrs, see \S \ref{discussion:origin}). Therefore, the rotational motion in NGC 4656 has not be able to homogenize the metal content in this galaxy, regardless the mechanism that produce the azimuthal chemical inhomogeneities.

Our oxygen abundance values for NGC 4656 obtained with the N2 index \citep{2013marino} are very similar to those found by \cite{2017zasov} doing the analysis with the same method, and are also similar to the values derived from \cite{2017zasov} with theoretical IZI method. However we found discrepancies with the abundances derived by previous authors when using the S method. %Although in this work we found that the values of oxygen abundance are lower in the northeast sector of NGC 4656 that in its optical center, in agreement with \cite{2017zasov}, this discrepancy is negligible considering the uncertainty in the estimates. 
In this paper we do not find strong drops in the metallicities along NGC 4656 (variations of the order of 0.3 dex or greater) as previously found by \cite{2017zasov}, who suggests that these drops can be associated with a gas accretion event \citep{2014sanchez,2015sanchez,2016ceverino}. 

Considering the results mentioned above, the following questions arise: why do metals are distributed almost homogeneously in the northeast sector of NGC 4656, and what is the reason for the lower metallicity in the southwest sector of the galaxy? The metallicity distribution and the presence of large H{\sc ii} regions in the northeast region of the galaxy suggest the existence of gas flows which are induced by the gravitational interaction with its companions (NGC 4631 and mainly with NGC 4656UV). %these flows in turn generate instabilities in the molecular clouds, which finally collapse forming stars.
In this way, in agreement with simulations simulations \citep[for example,][]{2010rupke_a} and the study of \cite{2015pilyuginb}, the gas mixture would produce a flattening in the metal distribution of interacting systems. These gas flows would also explain the non-rotational motions evidenced in the radial velocity maps (presented in Fig.\ref{vel_56y57}, \S \ref{sec:curves}). The southwest region of the galaxy, according to this scenario, would not be under the influence of strong gas flows as it happens in the northeast sector.

\subsection{Star formation}\label{formacion}

The SFRs$_{H\alpha}$ obtained for NGC 4656 and NGC 4656UV, taking into account their uncertainties, locate both galaxies in the star-forming sequence of galaxies considering the relations published by \cite{2012zahid} (log(M$_{*}$)$\sim$8.5-10.4) and \cite{2017mcgaugh} (log(M$_{*}$)$\sim$6.7-9.8) respectively, confirming their late-type nature.%, calibrated according to L$_{H\alpha}$.
We also found that the SFR$_{H\alpha}$ estimated for NGC 4656 is $\sim$3.5 times lower than the SFR$_{FUV}$ obtained by \cite{2012schech}, and for NGC 4656UV the SFR$_{H\alpha}$  is $\sim$4.5 times lower than SFR$_{FUV}$. Among the possible reasons of this discrepancy we can list: 1) variations in the IMF that cause a deficiency of massive ionizing stars in low mass galaxies, as suggested by \cite{2009lee}, who found that at low masses the luminosity in FUV is a better tracer of star formation compared to H$\alpha$, since the latter tends to underestimate the total SFR with respect to FUV, as lower-luminosity dwarf galaxies are examined, 2) the uncertainties caused by the indirect calibration of the Fabry-Perot cubes from the GMOS data(RMSE=1.32 $\times$10$^{-15}$erg s$^{-1}$cm$^{-2}$, see \S \ref{datos_fluxcalibration}).

On the other hand, we compared the SFRs$_{H\alpha}$ in the A, B and C zones of NGC 4656, obtaining the highest SFR$_{H\alpha}$ in the A sector ($\sim$2 times higher than in C) which indicates an enhancement in the star formation at the north region of this galaxy. It does not seem coincidence that this enhancement occurs closest to NGC 4656UV, where gas flows may be present. %which would also be under the influence of gas flows, with the southwest sector being the furthest from NGC 4656UV and which does not have an obvious influence of gas flows. %(see Fig. \ref{vel_56y57}, \S \ref{radia_distances}). 
On the other hand, the southwest region has the lowest star formation activity, which could be produced by a low H{\sc i} gas density, which is confirmed in the H{\sc i} maps presented by \cite{2012schech} (Fig. 2). This suggests that NGC 4656UV would be primarily responsible for increasing the star formation in the north of NGC 4656 (rather than NGC 4631), where the reasons are explained in the next subsection.

\subsection{Possible origin for NGC 4656UV}\label{discussion:origin}

% queda raro decir que no ha sido ampliamente estudiada y despues dices que dos trabajos se ahn enfocado en ella... Dos trabajos puede ser poco, pero podrias reescribir para evitar consufiones

The origin of NGC 4656UV is unclear.
%The TDG candidate, NGC 4656UV, has not been widely studied in literature, however 
Indeed, there are two studies that have proposed different scenarios regarding its nature. One of them is \cite{2012schech}, who proposes that NGC 4656UV has been formed from a recent interaction between NGC 4656 and NGC 4631 ($\sim$230 Myrs ago), mainly based on the small difference between both radial velocities (40 km s$^{-1}$), in the filamentary structures of H{\sc i} that arise from NGC 4631 to NGC 4656 \citep[found by][]{1978weli,1994rand}, and in the continuity of the the rotation gradient observed in the H{\sc i} maps of NGC 4656. However, for that kind of interactions we would expect to find more tidal characteristics in the H{\sc i} structure that surrounds NGC 4656 than those observed \citep{1978combes,2015martinez}. Also, there are no stellar tidal/bridges features connecting both galaxies or other optical evidence that indicate a previous baryonic interaction between NGC 4656 and NGC 4631 \citep{1983stayton}. For example, \cite{2015martinez} discovered a giant stellar tidal stream in the halo of NGC 4631, which extends between NGC 4631 and NGC 4656, however, these authors rule out that the origin of this stellar stream was produced by an interaction between NGC 4631 and NGC 4656 due to the orientation of this structure with respect to the orientations of the two galaxies, indicating that the most likely origin is the interaction between NGC 4631 and its satellites. On the other hand, the continuity of the rotation between NGC 4656 and NGC 4656UV found by \cite{2012schech} does not necessarily indicate that NGC 4656UV is part of the structure of NGC 4656. Similar is the case of the Magellanic Clouds, where both galaxies are inside a common  H{\sc i} envelope and are also connected in velocity \citep{2005bruns}, which is an evidence the the ongoing interaction between both galaxies. By fitting the SED of NGC 4656UV, \cite{2012schech} find that its metallicity is
approximately 10 times lower than the value reported for NGC 4656 \citep[(Fe/H)$\sim$-1.12,][]{2010mapelli}, thus relating the origin of NGC 4656UV with a possible star formation event associated with gas located in the outskirts of NGC 4656. However, in this work we do not find that difference. We have estimated an oxygen abundance for NGC 4656UV of 12+log(O/H)$\sim$8.2 which is similar to that obtained for NGC 4656, considering the uncertainties (12+log(O/H)$\sim$8.1). We have also estimated an abundance gradient for NGC 4656. If we extrapolate this gradient at large radius, we find that the oxygen abundance is 10 times lower than the central region at a radius of 300 kpc, which is not realistic (under the assumption that the abundance gradient can be extrapolated to large radius).

%from which it is deduced that, in order for the NGC 4656 disk to have at some radius a metallicity 10 times lower than in the center, this radius must be close to 300 kpc, which is not realistic (under the assumption that the abundance gradient can be extrapolated to large radius).

%Due to the low H$\alpha$ emission found in this work for NGC 4656UV (L$_{H\alpha}\sim$4$\times$10$^{38}$erg s$^{-1}$), and its high UV emission (L$_{FUV}\sim$7$\times$10$^{40}$erg s$^{-1}$), we agree with \cite{2012schech} on the fact that the last major star formation burst in this system occurred within the last $\sim$200-300 Myr.

Recently, \cite{2017zasov} used long-slit spectroscopy to obtain the kinematical parameters and the metallicity of NGC 4656UV and NGC 4656. They estimated the total dynamical mass of NGC 4656UV, concluding that this is a low surface brightness (LSB) dwarf galaxy, dominated by dark matter. They propose that the recent star formation in NGC 4656UV, responsible for its blue color, is probably the result of the accretion of low metallicity gas from NGC 4631. The same gas accretion could be feeding the northeast part of NGC 4656. As mentioned previously, based on our results of chemical abundances, in this work we do not find clear signs that indicate an accretion of gas from an external source.

Among our results, we found that the oxygen abundance for NGC 4656UV (12+log(O/H)$\sim$8.2) is very similar to that of NGC 4656 (12+log(O/H)$\sim$8.1) considering the uncertainties ($\sim$0.2 dex), which could indicate that both galaxies are 
forming stars from already enriched material, giving evidence in favor of a tidal origin for NGC 4656UV. However, it is also necessary to note that NGC 4656UV follows the mass-metallicity relation for dwarf galaxies of \cite{2015jimmy} (within the dispersion of the relation). In this sense, NGC 4656UV does not have a high metallicity for its mass, which characterizes the TDGs \citep{2003weil}.

Another possibility for the tidal origin of NGC 4656UV is that it could have been formed through the interaction between NGC 4656 and NGC 4657. However, as explained above, we found that NGC 4656 consists of only one galaxy. %in this way we did not find a companion galaxy to NGC 4656 with which it could have collided and subsequently formed a TDG.
Although in this work we can not rule out any scenario, and given the extreme complexity of the system, we propose that NGC 4656 and NCG 4656UV are a pair of interacting galaxies belonging to the NGC 4631 group, being the TDG candidate (NGC 4656UV) a  LSB galaxy \citep[with central magnitude 24 mag arsec$^{-1}$,][]{2017zasov} companion of NGC 4656, not of tidal origin but rather primordial.
An eventual interaction between NGC 4656 and NCG 4656UV could have triggered the star formation in NGC 4656UV around 200 Myrs ago, which is indicated by its UV emission. The weak  H$\alpha$ emission in NGC 4656UV suggests that this object is not experiencing an intense star-formation episode. %The above could be caused by the low density of neutral gas column in this system, where the values measured by \ citep {2012schech} in regions showing star formation (up to 1.6 M $ \ odot} $ pc $ - 2} $) are below the limit needed to trigger star formation \ citep [10 M $ \ odot} $ pc $ -2,] [] {2001martin}.

\section{Summary and conclusions} \label{sec:summary}

In this paper, we performed a spectroscopic and kinematic study of the interacting system NGC 4656/4656UV, through Gemini/GMOS data in multi-slit mode and Fabry-Perot data cubes, respectively, in order to find the possible origin of NGC 4656UV, a TDG candidate located in the northeast side of NGC 4656.
We obtained a low chemical abundance for NGC 4656UV (12+log(O/H)$\sim$8.2), which follows the mass-metallicity relation for normal dwarf galaxies of \cite{2015jimmy}. For NGC 4656 (12+log(O/H)$\sim$8.1), we estimated the oxygen abundance gradient of the northeast region, where most of the H{\sc ii} regions are located, obtaining a flat gradient  ($\beta$=-0.027$\pm$0.029 dex kpc$^{-1}$).  This is in agreement with the abundance gradients obtained in literature for other irregular dwarf galaxies \citep{1999hunter,2006vanzee,2009croxall}. %We compared the gradient obtained for NGC 4656 with that reported for NGC 55 (both galaxies with similar R$_{25}$ y M$_{*}$), and we note that both are flats within the uncertainties and also that in both galaxies the distribution of metals reach large projected galactocentric distances ($\sim$7 kpc) considering their low masses. 
In this work, we suggest that there are gas flows induced by the gravitational interaction between NGC 4656 and NGC 4631 and mainly with NGC 4656UV, which causes a mixture of gases and consequently a flattening of the gradient, consistent with simulations %****\textbf{otra cosa que no entiendo. Antes dices que el gradiente plano es normal en dwarf... por que la mezcla de gases aplanaria el gradiente en este caso, si pudiese ser plano solo por ser una galaxia dwarf? } 
\citep{2010rupke_a,2010montuori,2012torrey} and the study of \cite{2015pilyuginb} for irregular dwarf galaxies. 

%The FP kinematic data showed us the complexity of the system. Thanks to the spatial/spectral resolution of these data we were able to obtain new and detailed results about the behavior of the ionized gas in NGC 4656, of which there was no prior knowledge in the literature. Through the analysis of the H$\alpha$ emission profiles, we conclude that NGC 4656 consists of only one entity, and not 2 as was proposed by other authors \citep{1964vaucouleurs,1973nilson}. The velocity field in NGC 4656 suggests that this galaxy is sustained by rotation ((V$_{max}\sim$77 km s$^{-1}$), however at small scales we find non-rotational and counter-rotating motions, which agrees with the results of \cite{2012schech} for the H{\sc i} gas.\\

The FP kinematic data showed us the complexity of the system. Thanks to the spatial/spectral resolution of these data we were able to obtain new and detailed results about the behavior of the ionized gas in NGC 4656. %about which there was no prior knowledge in the literature. 
Through the analysis of the radial velocity profiles and by fitting a kinematic model of the observed FP velocity field, we conclude that NGC 4656 consists of only one entity, and not two as was proposed by other authors \citep{1964vaucouleurs,1973nilson}. From the kinematic model we estimated for NGC\,4656 a dynamical mass of $6.8^{1.8}_{-0.6}\times10^{9}$\,M$_{\sun}$
at a R=12.1\,kpc. The velocity field in NGC 4656 suggests that this galaxy is dominated by rotation (V$_{max}$/$\sigma\gtrsim$2.8), however at small scales we find non-rotational and counter-rotating motions, which agrees with the results of \cite{2012schech} for the H{\sc i} gas.

%In this work, thanks to the FP data and their indirect flux calibration through the GMOS spectra, we estimate SFRs for NGC 4656 and NGC 4656UV. Considering the distance to the system adopted in this work (5.1 Mpc), we obtained the values of  SFR$_{H\alpha}$=0.094 M$_{\odot}yr^{-1}$ and SFR$_{H\alpha}$=0.003 M$_{\odot}yr^{-1}$. We also assumed different distances to obtain new SFRs that were comparable with the results of other studies. For example, using a distance of 7.2$\pm$0.8 Mpc \citep{2005seth}, which was adopted by \cite{2012schech}, we obtained the value of SFR$_{H\alpha}$=0.186 M$_{\odot}yr^{-1}$ for NGC 4656 and SFR$_{H\alpha}$=0.006 M$_{\odot}yr^{-1}$   for NGC 4656UV, which differs from the estimates made by \cite{2012schech} (based on UV data). Our results are $\sim$3 and $\sim$4 times lower than the SFR$_{FUV}$  obtained by \cite{2012schech} for NGC 4656 and NGC 4656UV respectively. These discrepancies are in agreement with the study of \cite{2009lee}, who found that as dwarf galaxies of less luminosity are examined, H$\alpha$ tends to underestimate the total SFR with respect to FUV. Also, it is necessary to note that the flux calibration of the FP data cubes is not absolute, a factor that can influence the estimation of the luminosities.

Based on the results found in this work, supplemented with information from the literature, we suggest that NGC 4656 and NCG 4656UV are a pair of interacting galaxies belonging to the NGC 4631 group, with NGC 4656UV being a low surface brightness galaxy \citep[with central magnitude 24 mag arsec$^{-1}$]{2017zasov}, companion of NGC 4656, not of tidal origin but rather primordial. Its interaction with NGC 4656 triggered the star formation in NGC 4656UV, however the weak emission in H$\alpha$ suggests that this object is not experiencing an episode of intense star formation. Future studies of stellar populations based on IFS data are necessary to confirm this scenario. \\
%%%%%%%%%%%%%%%%%%%%%%%%%%%%%%%%%%%%%%%%%%%%%%%%%%
\section*{Acknowledgements}
 Based on observations taken at the Observatoire de Haute Provence (OHP) (France), operated by the French CNRS.  Based on data collected at the Gemini Observatory, which is operated by the Association of Universities for Research in Astronomy, Inc., under a cooperative agreement with the NSF on behalf of the Gemini partnership: the National Science Foundation (United States), the National Research Council (Canada), CONICYT (Chile), Ministerio de Ciencia, Tecnolog\'ia 
Innovaci\'on Productiva (Argentina), and Minist\'erio da Ci\^encia, Tecnologia e Inova\c c\~o (Brazil). 
NM-E acknowledges the financial support of the Direcci\'on de Investigaci\'on of the Universidad de La Serena, through a ```Concurso de Apoyo a Tesis 2015'', under contract PT1541. NM-E and FU-V acknowledges the financial support of the project CONICYT PAI 82140065.
 J.~A.~H.~J. thanks to Brazilian  institution CNPq for financial support through  postdoctoral fellowship (project 150237/2017-0).
 The authors thank Beno\^it Epinat, Michel Marcelin, Jean-Luc Gach \& Olivier Boissin for having made possible the OHP observations using the GHASP team instrument. In particular, we thank Beno\^it Epinat for optimizing the observational strategy to minimize ghosts due to parasitic reflections.


\bibliographystyle{mnras}
\bibliography{paper.bib}% if your bibtex file is called example.bib

%\begin{thebibliography}{99}
%\end{thebibliography}

%%%%%%%%%%%%%%%%%%%%%%%%%%%%%%%%%%%%%%%%%%%%%%%%%%

%%%%%%%%%%%%%%%%% APPENDICES %%%%%%%%%%%%%%%%%%%%%

%\appendix


%%%%%%%%%%%%%%%%%%%%%%%%%%%%%%%%%%%%%%%%%%%%%%%%%%

% Don't change these lines
\bsp	% typesetting comment
\label{lastpage}
\end{document}